\documentclass[aps, prd, superscriptaddress, nofootinbib, twocolumn, preprintnumbers]{revtex4-1}

\usepackage{bm,amssymb,slashed,graphicx,multirow,soul,mathtools,xspace,array}
\usepackage{enumitem}
\usepackage{makecell}
\usepackage{xcolor}

\definecolor{nicered}{rgb}{0.7,0.1,0.1}
\definecolor{nicegreen}{rgb}{0.1,0.5,0.1}
\definecolor{niceblue}{rgb}{0.1,0.1,0.5}

\usepackage[bookmarks=false]{hyperref}
\hypersetup{colorlinks,citecolor= nicegreen,linkcolor= niceblue,urlcolor= niceblue}

\newcommand{\nn}{\nonumber}

\def\spnt{1}
\if\spnt1
	\newcommand{\up}{+}
	\newcommand{\dn}{-}
\else
	\newcommand{\up}{2}
	\newcommand{\dn}{1}
\fi

\newcommand{\bbar}{\bar{b}}
\newcommand{\cbar}{\bar{c}}
\newcommand{\g}{\gamma}
\newcommand{\mn}{{\mu\nu}}
\newcommand{\GeV}{\text{GeV}}

\newcommand{\Lb}{\Lambda_b}
\newcommand{\Lc}{\Lambda_c}

\newcommand{\LcSp}{
	\mathchoice{\Lambda_c^*(\tfrac{1}{2}^{\!-}\!)}
			{\Lambda_c^*(1/2^-)}{\Lambda_c^*(1/2^-)}{\Lambda_c^*(1/2^-)}
}
\newcommand{\LcTn}{
	\mathchoice{\Lambda_c^*(\tfrac{3}{2}^{\!-}\!)}
		{\Lambda_c^*(3/2^-)}{\Lambda_c^*(3/2^-)}{\Lambda_c^*(3/2^-)}
}
\newcommand{\LSp}[1]{
	\mathchoice{\Lambda_{#1}^*(\tfrac{1}{2}^{\!-}\!)}
			{\Lambda_{#1}^*(1/2^-)}{\Lambda_{#1}^*(1/2^-)}{\Lambda_{#1}^*(1/2^-)}
}
\newcommand{\LTn}[1]{
	\mathchoice{\Lambda_{#1}^*(\tfrac{3}{2}^{\!-}\!)}
		{\Lambda_{#1}^*(3/2^-)}{\Lambda_{#1}^*(3/2^-)}{\Lambda_{#1}^*(3/2^-)}
}
\newcommand{\Lcs}{\Lambda_c^*}
\newcommand{\Psic}{{\Psi\kern-0.1em}_c}
\newcommand{\Psicbar}{{\overline{\phantom{h}}\kern-0.6em\Psi\kern-0.1em}_c}
\newcommand{\ccdot}{\!\cdot\!}

\newcommand{\aS}{\alpha_s}
\newcommand{\haS}{\hat{\alpha}_s}
\newcommand{\lqcd}{\Lambda_{\text{QCD}}}
\newcommand{\LamL}{\bar{\Lambda}_{\!\Lambda}}
\newcommand{\LamLp}{\bar{\Lambda}_{\!\Lambda}^\prime}
\newcommand{\ec}{\varepsilon_c}
\newcommand{\eb}{\varepsilon_b}
\newcommand{\pc}{\phi_c}
\newcommand{\pb}{\phi_b}
\newcommand{\hs}{\hat{\sigma}_1}
\newcommand{\hpc}{\hat{\phi}_c}
\newcommand{\hpb}{\hat{\phi}_b}

\newcommand{\alSL}{\alpha_L^S}
\newcommand{\alSR}{\alpha_R^S}
\newcommand{\alVL}{\alpha_L^V}
\newcommand{\alVR}{\alpha_R^V}
\newcommand{\alTL}{\alpha_L^T}
\newcommand{\alTR}{\alpha_R^T}
\newcommand{\beSL}{\beta_L^S}
\newcommand{\beSR}{\beta_R^S}
\newcommand{\beVL}{\beta_L^V}
\newcommand{\beVR}{\beta_R^V}
\newcommand{\beTL}{\beta_L^T}
\newcommand{\beTR}{\beta_R^T}
\newcommand{\rC}{r_{\Lambda^*}}

\newcommand{\rt}{r_l}
\newcommand{\mSqq}{\hat q^2}

\newcommand{\thtau}{\theta_{l}}
\newcommand{\phtau}{\phi_{l}}

\newcommand{\Wp}{w_+}
\newcommand{\Wm}{w_-}
\newcommand{\WpmP}{\Sigma_+}
\newcommand{\WpmM}{\Sigma_-}
\newcommand{\Ot}{\Omega_{\times}}
\newcommand{\Op}{\Omega_{+}}
\newcommand{\Oz}{\Omega_{0}}
\newcommand{\Rp}{R_{+}}
\newcommand{\Rm}{R_{-}}

\newcommand{\dS}{d_{S}}
\newcommand{\dP}{d_{P}}
\newcommand{\dV}[1]{d_{V#1}}
\newcommand{\dA}[1]{d_{A#1}}
\newcommand{\dT}[1]{d_{T#1}}
\newcommand{\lS}{l_{S}}
\newcommand{\lP}{l_{P}}
\newcommand{\lV}[1]{l_{V#1}}
\newcommand{\lA}[1]{l_{A#1}}
\newcommand{\lT}[1]{l_{T#1}}

\newcommand{\hdS}{\hat{d}_{S}}
\newcommand{\hdP}{\hat{d}_{P}}
\newcommand{\hdV}[1]{\hat{d}_{V#1}}
\newcommand{\hdA}[1]{\hat{d}_{A#1}}
\newcommand{\hdT}[1]{\hat{d}_{T#1}}
\newcommand{\hlS}{\hat{l}_{S}}
\newcommand{\hlP}{\hat{l}_{P}}
\newcommand{\hlV}[1]{\hat{l}_{V#1}}
\newcommand{\hlA}[1]{\hat{l}_{A#1}}
\newcommand{\hlT}[1]{\hat{l}_{T#1}}

\newcommand{\Cs}{C_{S}}
\newcommand{\Cps}{C_{P}}
\newcommand{\Cv}[1]{C_{V#1}}
\newcommand{\Ca}[1]{C_{A#1}}
\newcommand{\Ct}[1]{C_{T#1}}

\newcommand{\hammer}{\texttt{Hammer}\xspace}

\makeatletter
\g@addto@macro\bfseries{\boldmath}
\makeatother

\allowdisplaybreaks
\begin{document}

\preprint{CALT-2021-020}

\title{Form Factor Counting and HQET Matching for New Physics in $\Lambda_b\to\Lambda_c^*l\nu$}

\author{Michele Papucci}
\affiliation{Burke Institute for Theoretical Physics, 
California Institute of Technology, Pasadena, CA 91125, USA}

\author{Dean J.\ Robinson}
\affiliation{Ernest Orlando Lawrence Berkeley National Laboratory, 
University of California, Berkeley, CA 94720, USA}

\begin{abstract}
We calculate the $\Lambda_b \to \Lambda_c^*(2595) l \nu$ and $\Lambda_b \to \Lambda_c^*(2625) l \nu$ form factors and decay rates 
for all possible $b\to c l \bar\nu$ four-Fermi interactions in and beyond the Standard Model (SM), 
including nonzero charged lepton masses and terms up to order $\mathcal{O}(\alpha_s, 1/m_{c,b})$ in the heavy quark effective theory (HQET).
We point out a subtlety involving the overcompleteness of the representation of the spin-parity $1/2^+ \to 3/2^-$ antisymmetric tensor form factors, 
relevant also to other higher excited-state transitions,
and present a general method for the counting of the physical form factors for any hadronic transition matrix element and their matching onto HQET.
We perform a preliminary fit of a simple HQET-based parametrization of the $\Lambda_b \to \Lambda_c^*$ form factors at $\mathcal{O}(\alpha_s, 1/m_{c,b})$ 
to an existing quark model, providing preliminary predictions for the lepton universality ratios $R(\Lambda_c^*)$ beyond the SM.
Finally, we examine the possible incompatibility of recent lattice QCD results with expectations from the heavy-quark expansion and available experimental data.
\end{abstract}

\maketitle

\section{Introduction}
Exclusive $b \to c$ semileptonic decays to excited charm states play an intriguing dual role in probes of lepton flavor universality violation (LFUV), 
as well as extractions of the CKM matrix element $|V_{cb}|$. 
In the first instance, they contribute sizeable feed-down backgrounds to measurements from ground state decays,
leading to systematic uncertainties that are of comparable size to the statistical ones in the current $3\sigma$~\cite{Amhis:2019ckw} (or more~\cite{Bernlochner:2021vlv}) 
evidence for LFUV seen in $B \to D^{(*)} \tau \nu$ decays.
In the second instance, they are signal modes in their own right, through which LFUV or $|V_{cb}|$ might be directly measured.

In light of the large anticipated datasets from LHCb and Belle II, reducing systematic uncertainties from excited state decays to a sufficiently low level---roughly 
the percent level---will be essential for establishing conclusive tensions with the Standard Model (SM) (see e.g. Ref.~\cite{Bernlochner:2021vlv} for a review).
Further, identifying what New Physics (NP) operators are \emph{compatible} with (future) $b \to c \tau \nu$ data for not only ground state decays but also excited state decays
requires good control of the description of the latter in the SM and beyond. 
In the case of $B \to D^{(*)} \tau \nu$ decays, 
this mandates precision NP predictions for the semitauonic decays to orbitally excited charm meson final states, i.e. $B \to D^{**} l \nu$~\cite{Bernlochner:2016bci,Bernlochner:2017jxt} ($l = e$, $\mu$ or $\tau$).
In the case of baryons, especially for studying NP effects in $\Lb \to \Lc \tau\nu$ decays and the relevant 
form-factors~\cite{Detmold:2015aaa, Dutta:2015ueb, Datta:2017aue, Li:2016pdv, Bernlochner:2018kxh, Bernlochner:2018bfn,Hu:2020axt} 
(see also Refs~\cite{Azizi:2018axf,Faustov:2016pal,Gutsche:2014zna,Ke:2007tg,Mannel:2015osa})
one requires NP predictions for the $\Lb \to \Lcs l \nu$ excited state decays,
where $\Lcs$ is the $s_\ell^P = 1^-$ Heavy Quark Symmetry (HQS) doublet composed of the $\LcSp$ and $\LcTn$.

For the vector and axial-vector $\Lb \to \Lcs$ currents only, Ref.~\cite{Leibovich:1997az} long ago studied the $s_\ell^P = 1^-$ Heavy Quark Effective Theory (HQET)
to $\mathcal{O}(1/m_{c,b})$ in the heavy quark (HQ) expansion.
Subsequently, Ref.~\cite{Boer:2018vpx} recently included the $\mathcal{O}(\aS)$ radiative corrections for the SM currents
(though dropping the contributions from $1/m_{c,b}$ chromomagnetic corrections).
Because NP predictions to $\mathcal{O}(1/m_{c,b},\aS)$ in the HQET for $\Lb \to \Lcs l \nu$ are not present in the Literature, 
we therefore provide them, 
based on a treatment similar to that developed in Refs.~\cite{Bernlochner:2018kxh,Bernlochner:2018bfn} for $\Lb \to \Lc \tau \nu$ 
and Refs.~\cite{Bernlochner:2016bci,Bernlochner:2017jxt} for $B \to D^{**}\tau\nu$.
The constrained structure of the $s_\ell^P = 0^+$ HQET for $\Lb \to \Lc l \nu$ 
has already permitted the size of the contributions at $\mathcal{O}(1/m_{c}^2)$ in that decay to be extracted 
from combined fits to data plus lattice QCD (LQCD) calculations~\cite{Bernlochner:2018kxh}. 
These higher-order corrections were found to be compatible with a well-behaved HQ expansion.
It is therefore not unreasonable to expect a similarly well-behaved HQ expansion for $\Lb \to \Lcs$ transitions.

However, unlike in the study of $B \to D^{(*,**)}l\nu$ decays or ground state baryon decays, 
a subtlety arises for the $\LcTn$, when one applies the usual HQET matching approach: 
We observe that the form factor representation of the  $\Lb \to \LcTn$ matrix element for the antisymmetric tensor current, $\cbar \sigma_{\mn} b$, 
is overcomplete with respect to the basis of physical amplitudes.
That is, there exists a linear combination of terms in the form factor representation of the matrix element---a kernel---that 
does not appear in any physical amplitude. 
Preserving such redundancy in the matrix elements can be essential for keeping the HQET symmetries manifest at any order in the HQS-breaking expansion while performing consistent calculations. 
Moreover, HQET generates relations between the form factors that must be obeyed order-by-order in $1/m$, 
such that this kernel can be removed only after HQET matching is imposed on the tensor current.

Anticipating this subtlety and its relevance also to other higher excited state decays, we first present in Section~\ref{sec:cffs} a general discussion of the 
counting of the physical form factors for any hadronic transition matrix element,
showing how to determine the kernel with respect to the physical amplitudes, and examining the role of HQET matching.
We then proceed in Sec.~\ref{sec:match} to apply these results to $\Lb \to \Lcs l \nu$ decays, matching onto HQET at $\mathcal{O}(1/m_{c,b}, \alpha_s)$.
We thus derive the HQET expansion of the form factors at this order for all possible NP currents.

Currently no experimental data (neither at nor beyond zero-recoil) is available for $\Lb \to \Lcs l \nu$,
with the notable exception that there exist measurements of their relative branching fractions 
compared to $\Lb \to \Lc l \nu$~\cite{Aaltonen:2008eu}.
(There is, in addition, ongoing work at LHCb, with preliminary results for differential spectra presented in Ref.~\cite{Lupato:2315592}. 
These results are not yet published because of challenging charm backgrounds.)
The ratio of these branching fractions are compatible with expectations from the HQ expansion.
As a stopgap measure, in Sec.~\ref{sec:fits} we therefore fit the HQ expansion of the form factors at order $\mathcal{O}(1/m_{c,b}, \alpha_s)$ over the full recoil range
to the predictions of the PCR parametrization~\cite{Pervin:2005ve}.
This parametrization is based on a constituent quark model approach similar to ISGW2~\cite{Scora:1995ty,Isgur:1988gb};
the covariant confined quark model~\cite{Gutsche:2017wag,Gutsche:2018nks} and relativistic quark models~\cite{Becirevic:2020nmb}
have also been recently applied to generate $\Lb \to \Lcs l \nu$ predictions.
Using our PCR-based fit, we generate preliminary predictions for a simple parametrization of the Isgur-Wise functions up to linear order in the recoil parameter, $w-1$, 
that might be plausibly used as central values for a future data-driven fit.
We further examine the behavior of the lepton flavor universality ratios $R(\Lcs)$ for both the SM and in the presence of various NP currents.
In addition, in an Appendix we present the NP helicity amplitudes for $\Lb \to \Lcs l \nu$, 
to be used in the \hammer library~\cite{Bernlochner:2020tfi,bernlochner_florian_urs_2020_3993770}.

Very recently, new LQCD results~\cite{Meinel:2021rbm} have also become available for $\Lb \to \Lcs l \nu$, 
in the near-zero recoil regime $w \le 1.05$. 
We examine the compatibility of these lattice results with HQET, 
fitting the HQET expansion for the SM and NP form factors at $\mathcal{O}(1/m_{c,b},\aS)$ to the LQCD predictions. 
As remarked in passing in Ref.~\cite{Meinel:2021rbm} regarding the SM form factors at $\mathcal{O}(1/m_{c,b})$,
we similarly find that this fit appears to be very poor.
We briefly characterize the implied degree of HQS breaking
by modifying the fit to include additional HQS-breaking parameters.
We find this modified fit implies $\mathcal{O}(1)$ violations of HQS at $\mathcal{O}(1/m_{c,b}, \alpha_s)$: 
A surprising result, demanding further study, given the well-behaved convergent behavior of the HQ expansion in $\Lb \to \Lc l \nu$~\cite{Bernlochner:2018kxh}.
Curiously, we point out that while HQS expectations are compatible with the measured ratio of $\Lb \to \Lcs l \nu$ branching fractions~\cite{Aaltonen:2008eu},
the modified fit to the LQCD results is not.

\section{Counting form factors}
\label{sec:cffs}
\subsection{General method}
In a similar spirit to the discussion of Ref~\cite{Ji:2000id}, the counting of physical form factors for a hadronic transition $H_1 \to H_2$, mediated by an operator $\mathcal{O}$, 
can be systematically implemented by the counting of partial wave amplitudes in the crossed process: $\bar{H}_1 H_2$ pair-production by $\mathcal{O}$ from the vacuum.
For any particular operator $\mathcal{O}$, one decomposes it into a set of operators of definite angular momentum-parity, $\mathcal{O}_{J^P}$,
and then counts the number of partial waves of the $\bar{H}_1 H_2$ final state that may project onto each operator $\mathcal{O}_{J^P} \subset \mathcal{O}$.\footnote{
If $\mathcal{O}$ and $\bar{H}_1 H_2$ have also a well-defined charge-conjugation parity, then one matches with respect to $J^{PC}$.}
That is, the full matrix element 
\begin{equation}
	\big\langle \bar{H}_1 H_2 \big| \mathcal{O} \big| 0 \big\rangle = \sum_{J^P(L)} \! F_{J^P(L)} \big\langle (\bar{H}_1 H_2)_{J^P(L)} \big| \mathcal{O}_{J^P} \big| 0 \big\rangle\,,
\end{equation}
in which $(\bar{H}_1 H_2)_{J^P(L)}$ denotes a partial wave state with total (orbital) angular momentum-parity $J$ ($L$) and parity $P$, and $F_{J^P(L)}$ is its form factor.
As usual, each partial wave amplitude $\big\langle (\bar{H}_1 H_2)_{J^P(L)} \big| \mathcal{O}_{J^P} \big| 0 \big\rangle$ 
can in turn be represented as a tensor product of the external momenta, polarizations and/or spins, as appropriate to the external states.

An explicit algorithm to count such amplitudes proceeds as follows:
\begin{enumerate}[wide, labelwidth=!, labelindent=0pt, label = \quad \textbf{(\roman*)}, noitemsep, topsep =0pt]
	\item Write down the $J^P$ quantum numbers for the decomposition of the operator $\mathcal{O}$.
	\item For each choice of the orbital angular momentum, $L = 0$, $1$, $2$, \ldots, write down the possible $J^P$ values of the $\bar{H}_1 H_2$ partial wave decomposition,
		until $L$ is sufficiently high that one can no longer match onto any $\mathcal{O}_{J^P} \subset \mathcal{O}$.
	\item Count the number of $(\bar{H}_1 H_2)_{J^P(L)}$ partial waves that match onto each $\mathcal{O}_{J^P}$ and sum over $J^P$. This counts the number of independent amplitudes, 
	and thus the number of possible form factors, in the $H_1 \to H_2$ process mediated by $\mathcal{O}$.
\end{enumerate}

For $b \to c$ transitions, we are interested in operators of the form $\mathcal{O} = \cbar\, \Gamma\, b$, 
where $\Gamma = 1$, $\g^5$, $\g^\mu$, $\g^\mu\g^5$ or $\sigma^{\mn} \equiv (i/2)[\gamma^\mu, \gamma^\nu]$, 
for scalar (S), pseudoscalar (P), vector (V), axial-vector (A), and tensor (T) operators, respectively.
Their angular momentum-parity decompositions are
\begin{subequations}
\label{eqn:JPO}
\begin{align}
	J^P(\mathcal{O}_S) & = 0^+\,,  & J^P(\mathcal{O}_P) & = 0^-\,, \\
	J^P(\mathcal{O}_V) & = 0^+ \oplus 1^- \,, & J^P(\mathcal{O}_A) & = 0^- \oplus 1^+ \,, \\
	J^P(\mathcal{O}_T) & = 1^+ \oplus 1^-\,.
\end{align}
\end{subequations}
Thus we only need consider partial waves up to $J = 1$.

We employ the conventions $\text{Tr}[\g^\mu\g^\nu\g^\rho\g^\sigma\g^5] = -4i\epsilon^{\mu\nu\rho\sigma}$, so that the tensor-pseudotensor operator identity
\begin{equation}
	\label{eqn:tenaxten}
	\sigma^{\mu\nu} \g^5 \equiv +(i/2)\epsilon^{\mu \nu \rho \sigma} \sigma_{\rho \sigma}\,,
\end{equation}
matching the choices in Refs.~\cite{Bernlochner:2018bfn,Bernlochner:2018kxh} for $\Lb \to \Lc l \nu$ 
and the standard choice for excited charm mesons~\cite{Leibovich:1997tu, Leibovich:1997em, Bernlochner:2016bci,Bernlochner:2017jxt}. 
(For $\bar{B} \to D^{(*)}l\bar\nu$, the Literature typically chooses $\text{Tr}[\g^\mu\g^\nu\g^\rho\g^\sigma\g^5]= +4i \epsilon^{\mu\nu\rho\sigma}$, 
so that instead $\sigma^{\mu\nu} \g^5 \equiv -(i/2)\epsilon^{\mu \nu \rho \sigma} \sigma_{\rho \sigma}$.)
Thus, for any representation of the tensor current matrix element, the representation for the pseudotensor current follows by applying the identity~\eqref{eqn:tenaxten}, 
or the underlying Chisholm identity $\varepsilon_{\mu\nu\rho\alpha} \g^\alpha  \equiv -i/2 \big[\g_\mu\g_\nu \g_\rho - \g_\rho\g_\nu\g_\mu\big]\g^5$.

As an initial counting example, the $B \to D^*$ transition corresponds to the crossed process pair-production of a state with intrinsic spin-parity $s^{P^\prime}  = 0^- \otimes 1^- = 1^+$.
Further tensoring with a spatial partial wave with orbital angular momentum $L = 0$, $1$, $2$, \ldots and parity $(-1)^L$, one then has states
\begin{align}
	J^P(L=0) & = 1^+ \otimes 0^+  = 1^+\,, \label{eqn:JPBDs}\\
	J^P(L=1) & = 1^+ \otimes 1^-  = 0^- \oplus 1^- \oplus 2^-\,,\nn \\
	J^P(L=2) & = 1^+ \otimes 2^+  = 1^+ \oplus 2^+ \oplus 3^+\,, \nn
\end{align}
and no partial wave with $L > 2$ contains $J \le 1$.
Matching Eqs.~\eqref{eqn:JPBDs} onto~\eqref{eqn:JPO}, one may immediately read off the number of form factors, $N$, for each operator, viz:
$N(\mathcal{O}_S) = 0$, $N(\mathcal{O}_P) = 1$, $N(\mathcal{O}_V) = 1$, $N(\mathcal{O}_A) = 3$, $N(\mathcal{O}_T) = 3$. 
This result is exactly as would be expected from the more familiar process of writing down representations of the form factors 
explicitly in terms of the $D^*$ polarization and the $B$ and $D^*$ momenta, 
and imposing angular momentum and parity conservation.

\subsection{\texorpdfstring{$\Lb \to \LcSp$}{FFLbLcSp}}

Turning to $\Lb \to \LcSp$, this corresponds to pair-production of state with intrinsic spin-parity $s^{P^\prime}  = \frac{1}{2}^+\otimes\frac{1}{2}^- = 0^- \oplus 1^-$.
Further taking a tensor product with spatial partial waves, one has
\begin{align}
	J^P(L=0) & = (0^- \oplus 1^-) \otimes 0^+ 	 = 0^- \oplus 1^-\,, \label{eqn:JPLbLcSp}\\
	J^P(L=1) & = (0^- \oplus 1^-) \otimes 1^- 	 = 1^+   \oplus 0^+ \oplus 1^+ \oplus 2^+\,, \nn \\
	J^P(L=2) & = (0^- \oplus 1^-) \otimes 2^+  = 2^-  \oplus 1^- \oplus 2^- \oplus 3^-\,, \nn
\end{align}
and no partial wave with $L > 2$ contains $J \le 1$.
This time one counts $N(\mathcal{O}_S) = 1$, $N(\mathcal{O}_P) = 1$, $N(\mathcal{O}_V) = 3$, $N(\mathcal{O}_A) = 3$, $N(\mathcal{O}_T) = 4$. 

By comparison, one may define the form factors via an explicit representation of the matrix elements in terms of spinors and momenta,
writing down all possible independent combinations subject to angular momentum and parity conservation.
To do this, we follow the (classic) notation of Ref.~\cite{Leibovich:1997az} for the vector and axial-vector currents, extending them to $\mathcal{O}_{S,P,T}$. 
This yields
\begin{align}
	\label{eqn:HQETffdefSp}
	\langle \LcSp | \bar c\, b |\Lb\rangle &= -\dS\, \bar u_c \g_5 u_b\,, \\
	\langle \LcSp | \bar c \g_5 b |\Lb \rangle &= -\dP \, \bar u_c\, u_b\,, \nn\\*
	\langle \LcSp | \bar c\g_\mu b |\Lb \rangle
		&= \bar u_c \big[ \dV1 \g_\mu + \dV2 v_\mu + \dV3 v'_\mu \big]\g_5 u_b\,, \nn\\
	\langle \LcSp | \bar c\g_\mu\g_5 b |\Lb \rangle
 		 &= \bar u_c \big[ \dA1 \g_\mu + \dA2 v_\mu + \dA3 v'_\mu \big] u_b\,, \nn \\
	\langle \LcSp | \bar c\, \sigma_{\mn}\, b |\Lb \rangle
  		&= -\bar u_c \big[ \dT1\, \sigma_{\mn} + i\, \dT2 v_{[\mu} \g_{\nu]}   \nn \\
		& \quad + i\, \dT3 v'_{[\mu} \g^{\phantom{\prime}}_{\nu]} + i\, \dT4 v^{\phantom{\prime}}_{[\mu} v'_{\nu]} \big] \g_5 u_b\,. \nn 	 
\end{align}
where $\Lb(p,s)$ and $\LcSp(p', s')$ are represented by spinors $u_b(p,s)$ and $\bar u_c(p',s')$, respectively,
with momenta  $p = m_{\Lb}v$ and $p' = m_{\Lcs}v'$.
The form factors $d_X$ are
functions of 
\begin{equation}
	w = v \cdot v' = (m_{\Lb}^2 + m_{\Lcs}^2 - q^2)/(2 m_{\Lb} m_{\Lcs})\,, 
\end{equation}
and the spinors are normalized to $\bar u u= 2m$. 
We use the antisymmetrized index notation $x_{[\mu} \ldots y_{\nu]} \equiv x_\mu \ldots y_\nu - x_\nu \ldots y_\mu$.
Happily, the number of form factors deduced from the explicit representation~\eqref{eqn:HQETffdefSp}
is exactly the same as from counting partial waves.

\subsection{\texorpdfstring{$\Lb \to \LcTn$}{FFLbLcTn}}
We can now turn to the $\Lb \to \LcTn$ process, corresponding to the pair production of a state with intrinsic spin-parity $s^{P^\prime}  = \frac{1}{2}^+\otimes\frac{3}{2}^- = 1^- \oplus 2^-$,
such that
\begin{align}
	J^P(L=0) 	& = 1^- \oplus 2^-\,, \label{eqn:JPLbLcTn}\\
	J^P(L=1) 	& =  0^+ \oplus 1^+ \oplus 2^+ \nn \\
			& \quad \oplus 1^+ \oplus 2^+ \oplus 3^+\,,  \nn \\
	J^P(L=2) 	& =  1^- \oplus 2^- \oplus 3^- \nn \\
			& \quad \oplus 0^- \oplus 1^- \oplus 2^- \oplus 3^- \oplus 4^-\,, \nn \\
	J^P(L=3) 	& =  2^+ \oplus 3^+ \oplus 4^+ \nn \\
			& \quad \oplus 1^+ \oplus 2^+ \oplus 3^+ \oplus 4^+ \oplus 5^+\,. \nn 		
\end{align}
This time no partial wave with $L > 3$ contains $J \le 1$.
The form factor counting is $N(\mathcal{O}_S) = 1$, $N(\mathcal{O}_P) = 1$, $N(\mathcal{O}_V) = 4$, $N(\mathcal{O}_A) = 4$, $N(\mathcal{O}_T) = 6$. 

For the (pseudo)scalar and (axial-)vector, an explicit representation corresponding to this counting follows straightforwardly.
First, we represent the charmed spin-$3/2$ state by a Rarita-Schwinger tensor~\cite{PhysRev.60.61}, $\Psic^\mu(p',s')$, 
satisfying the usual transversity and projective conditions $v' \ccdot \Psic = 0$ and $\gamma \ccdot \Psic = 0$, 
and normalized such that $\Psicbar \ccdot \Psic = 2m_{\Lcs}$.
Note also $\slashed{v}^\prime \Psic = \Psic$.
Then, we have explicit form factor representations
\begin{align}
	\label{eqn:HQETffdefTn}
	\langle \LcTn | \bar c\, b |\Lb\rangle &= \lS\, v \ccdot \Psicbar  u_b\,, \\
	\langle \LcTn | \bar c \g_5 b |\Lb \rangle &= \lP \, v\ccdot \Psicbar \g_5 u_b\,, \nn\\*
	\langle \LcTn | \bar c\g_\mu b |\Lb \rangle
		&= v\ccdot \Psicbar \big[ \lV1 \g_\mu + \lV2 v_\mu + \lV3 v'_\mu \big] u_b \nn \\
		& \qquad + \lV4 {\Psicbar}_\mu u_b \,, \nn \\
	\langle \LcTn | \bar c\g_\mu\g_5 b |\Lb \rangle
 		 &= v\ccdot \Psicbar \big[ \lA1 \g_\mu + \lA2 v_\mu + \lA3 v'_\mu \big] \g_5 u_b \nn \\ 
		 & \qquad + \lA4 {\Psicbar}_\mu \g_5 u_b \,, \nn 
\end{align}
where the form factors $l_{X}$ are functions of the recoil parameter $w$, 
and we have followed the notation of Ref.~\cite{Leibovich:1997az} for the vector and axial-vector currents.
The number of form factors follows the counting from Eqs.~\eqref{eqn:JPLbLcTn}. 
The Rarita-Schwinger tensor itself can be represented in terms of spinors and polarizations (see e.g. Ref.~\cite{Boer:2018vpx}) as
\begin{equation}
	\Psic^\mu(p',s') = \Big[\varepsilon_c^\mu - \frac{1}{3}\big(\g^\mu + v^{\prime \mu}\big)\slashed{\varepsilon}_c\Big]u_{c}\,,
\end{equation}
in which $\varepsilon_c$ is the polarization with velocity $v'$ and helicity $\lambda' = \pm,0$.
Alternatively, it can be written as a Clebsch-Gordan decomposition with highest weight state $\varepsilon_c^+u_c^+$, 
using a spinor-helicity formalism representation for $\varepsilon_c$ and $u_c$.
Appropriate phase convention choices enforce the above transversity, projective and normalization conditions.

A complication, however, arises for the tensor current, for which an explicit representation is\footnote{
We have chosen a convention in which the form factor for the $v^{\phantom{\prime}}_{[\mu} v'_{\nu]}$ term is numbered last, i.e. $\lT7$, rather than $\lT4$.}
\begin{align}
	\langle \LcTn | \bar c\, \sigma_{\mn}\, b |\Lb \rangle
  		&= v\ccdot \Psicbar \big[ \lT1\, \sigma_{\mn} + i\, \lT2 v_{[\mu} \g_{\nu]}  \label{eqn:HQETffdefTnT} \\
		& \qquad + i\, \lT3 v'_{[\mu} \g^{\phantom{\prime}}_{\nu]} + i\, \lT7 v^{\phantom{\prime}}_{[\mu} v'_{\nu]} \big]u_b  \nn\\
		&+ i\, {\Psicbar}_{[\mu} \big[ \lT4\, \g^{\phantom{\prime}}_{\nu]} + \lT5 v^{\phantom{\prime}}_{\nu]} + \lT6 v'_{\nu]} \big] u_b \,. \nn
\end{align}
In Eq.~\eqref{eqn:HQETffdefTnT} there appears to be seven terms---and thus seven form factors---in 
the representation of the tensor current matrix element,
rather than six as derived from Eqs.~\eqref{eqn:JPLbLcTn} above. 
These terms are independent, such that one (or more) cannot be eliminated simply by applying equations of motion, or the transversity conditions.

\subsection{Why does 6 hate 7?}

The resolution to this puzzle begins by noting 
that locality and unitarity allow us to describe the  $\Lb \to \LcTn$ decay, mediated by $\cbar \sigma^{\mn} b$, 
via an on-shell amplitude between $\Lb$, $\LcTn$ and a continuum of fictitious massive particles with momentum 
\begin{equation}
	q = p - p'\,,
\end{equation}
mass $\sqrt{q^2}$,
and quantum numbers $J^P = 1^- \oplus 1^+$~\cite{Arkani-Hamed:2017jhn}\footnote{This is possible in the timelike region where $q^2 >0$ and can be extended elsewhere via analytic continuation.}.
Contracting the $\langle \LcTn | \bar c\, \sigma_{\mn}\, b |\Lb \rangle$ matrix element with a suitable polarization tensor representation of the $1^- \oplus 1^+$ states,
the generated on-shell amplitudes can be expressed in the form 
\begin{equation}
	\mathcal{A}_{\alpha} = \sum_i \mathcal{M}_{\alpha i} \, \lT{i}
\end{equation}
where $\alpha$ indexes a basis of physical amplitudes that describe the $\Lb \to \LcTn + q$ transition.
The key claim is that $\mathcal{M}$ must have a non-trivial kernel of dimension one in $\lT{i}$ space.
That is, there is a particular linear combination of terms in Eq.~\eqref{eqn:HQETffdefTnT} that cannot appear in a physical $\Lb \to \LcTn + q$ amplitude,
allowing one of the seven apparent form factors to be redefined away.

To build suitable antisymmetric tensor representations for (the polarizations of) the $1^- \oplus 1^+$ states, 
one may start from the conventional polarization vectors $\varepsilon(q)^\lambda_\mu$, with $\lambda = \pm,0$, 
satisfying transversity $\varepsilon(q) \cdot q = 0$ and space-like normalization $\varepsilon(q)^{\kappa*}\cdot\varepsilon(q)^\lambda = -\delta^{\kappa\lambda}$. 
Then one may proceed to construct the bilinears $\varepsilon(q)^{\lambda}_\mu \varepsilon(q)^{\kappa}_\nu$ and  $\varepsilon(q)^{\lambda}_{\mu} q^{\phantom{\kappa}}_{\nu}$. 
However these bilinears are not independent, as can be easily shown by representing both the polarization and momentum in the $SL(2,\mathbb{C})$ covering space 
by Weyl spinors,\footnote{The entire procedure is in fact much more straightforward using the spinor-helicity formalism, as shown in Appendix~\ref{app:helicity}. 
We have presented it here using 4-vectors in order not to change notation.}
from which one finds
\begin{align}
	\sigma^{\mn} \varepsilon(q)^{+}_\mu \varepsilon(q)^{0}_\nu & = -i\sigma^{\mn}\g^5 \varepsilon(q)^{+}_\mu q_\nu/\sqrt{q^2}\,,\nn\\
	\sigma^{\mn} \varepsilon(q)^{+}_\mu \varepsilon(q)^{-}_\nu & = -i\sigma^{\mn}\g^5 \varepsilon(q)^{0}_\mu q_\nu/\sqrt{q^2}\,,\nn\\
	\sigma^{\mn} \varepsilon(q)^{-}_\mu \varepsilon(q)^{0}_\nu & = +i\sigma^{\mn}\g^5 \varepsilon(q)^{-}_\mu q_\nu/\sqrt{q^2}\,, \label{eqn:epsqidentity}
\end{align}
(up to phase conventions for the polarizations).
Therefore, it is sufficient to consider only either $\varepsilon_\mu \varepsilon_\nu$ or $\varepsilon_\mu q_\nu$ when contracting against both the tensor and pseudotensor currents. 
In the following we will use the $\varepsilon_\mu q_\nu$ representation, as it is more convenient.

Contracting the tensor and pseudotensor currents with $q_\nu = m_{\Lb} v_\nu - m_{\Lcs} v'_\nu$, 
and applying equations of motion and transversity to eliminate all linearly dependent amplitudes, 
one may compute $\mathcal{M}_{\alpha i}$. 
One finds that $\{\lT{i}\} = (1,0,-1,w+1,-1,-1,0)$ is the kernel $\mathcal{M}$. 
From Eq.~\eqref{eqn:HQETffdefTnT}, this  corresponds to the combination of operators
\begin{align}
	\label{eqn:Kexp}
	\mathcal{K}^{\mu\nu} & =  \lambda_T\Big(v\ccdot \Psicbar\big[\sigma_{\mn} - i v'_{[\mu} \g^{\phantom{\prime}}_{\nu]}\big]u_b \nn \\
	\quad & + i\, {\Psicbar}_{[\mu} \big[(w+1) \g^{\phantom{\prime}}_{\nu]} -  v^{\phantom{\prime}}_{\nu]} -  v'_{\nu]} \big] u_b\Big)\,,
\end{align}
where the $\lambda_T$ is arbitrary. 
That is, this combination of operators does not contribute to any physical amplitude.
Put a different way,
adding $\mathcal{K}^{\mu\nu}$ to Eq.~\eqref{eqn:HQETffdefTnT}, the latter can be rewritten with the replacements
\begin{align}
	\lT1 & \to \lT1 + \lambda_T \,, & \lT3 & \to \lT3 - \lambda_T\,, \nn \\
	\lT4 & \to \lT4 + (w+1)\lambda_T \,, \nn \\
	\lT5 & \to \lT5 - \lambda_T  \,, & \lT6 & \to \lT6 - \lambda_T\,, \label{eqn:lTredef}
\end{align}
without affecting any physical amplitude.
For instance, one could choose $\lambda_T = \lT6$, 
thereby eliminating the $\lT6$ term from Eq.~\eqref{eqn:HQETffdefTnT}, and leaving a basis of only six physical form factors
(after suitable redefinitions) as expected from the partial wave counting.

However, as we shall see in the next Section, in order to be consistent with HQET at a particular order, 
care needs to be taken to \emph{first} match the tensor current matrix elements onto HQET, before eliminating terms.
This is not dissimilar to the case of the renormalization group evolution in the SM effective field theory: 
To consistently capture the operator mixing under running, one has to use an overcomplete operator basis 
and then match it to the basis of choice using kernels generated by equations of motion and integration by parts~\cite{Jenkins:2013zja}.
For instance, in Ref.~\cite{Mott:2011cx} the preemptive omission of the $\lT6$ term obscures the presence of the kernel in this procedure; 
we shall see below this omission happens to be inconsistent beyond leading order in HQET.

Before proceeding, some further comments are in order. 
Noting the relations~\eqref{eqn:epsqidentity}, to determine a basis of physical form factors, 
one could have instead defined them with respect to $\Gamma = \sigma_{\mn}q^\nu$ and $\sigma_{\mn}\g_5 q^\nu$, via
\begin{subequations}
\label{eqn:axqtenaxq}
\begin{align}
	\langle \LcTn | \bar c\sigma_\mn q^\nu b |\Lb \rangle
		&= v\ccdot \Psicbar \big[ F^T_1 \g_\mu + F^T_2 v_\mu + F^T_3 v'_\mu \big] u_b \nn \\
		& \qquad + F^T_4 {\Psicbar}_\mu u_b \,,  \\
	\langle \LcTn | \bar c\sigma_\mn q^\nu \g_5 b |\Lb \rangle
		&= v\ccdot \Psicbar \big[ G^T_1 \g_\mu + G^T_2 v_\mu   \\
		& \qquad + G^T_3 v'_\mu \big] \g_5u_b  + G^T_4 {\Psicbar}_\mu \g_5 u_b \,, \nn
\end{align}
\end{subequations}
as done in Refs.~\cite{Mott:2011cx,Meinel:2020owd} in the context of rare dileptonic decays $\Lb \to \Lambda^* \ell \ell$.
The identity~\eqref{eqn:tenaxten} applied to the underlying tensor structure~\eqref{eqn:HQETffdefTnT} 
would then allow one to show that these eight form factors
may be expressed in terms of just six linear combinations of $\lT{i}$, with the same kernel.
Note that the kernel itself satisfies both $\mathcal{K}^{\mu\nu}v_\nu = 0$ and $\mathcal{K}^{\mu\nu}v'_\nu = 0$ separately,
not just $\mathcal{K}^{\mu\nu}q_\nu = 0$. 


Alternatively, one could have imposed the constraint that the matrix elements~\eqref{eqn:axqtenaxq} must each vanish under further contraction with $q^\mu$, 
to directly remove one each of the $F^T_i$ and $G^T_i$, leaving a total of six form factors.
As done in Refs.~\cite{Descotes-Genon:2019dbw,Meinel:2020owd}, 
these six can be re-expressed in a so-called helicity basis, comprising three form factors for each matrix element.
Matching further onto the $\Gamma = \sigma_{\mn}q^\nu$ and $\sigma_{\mn}\g_5 q^\nu$ HQET matrix elements will yield consistent results,
because the HQET traces implicitly impose the identity~\eqref{eqn:tenaxten} or equivalent.
The helicity basis, however, is not a natural choice for straightforward HQS interpretations of the structure of the matrix elements.

Further, including $q = m_{\Lb} v_\nu - m_{\Lcs} v'_\nu$ in the definition of the (pseudo)tensor current itself in Eq.~\eqref{eqn:axqtenaxq} 
introduces hadron mass terms into the HQET traces, 
and it does not respect the reparametrization invariance of HQET: The momentum difference of the heavy quarks need not be $q$.
These hadron mass terms correspond to heavy quark mass terms plus HQS-breaking higher-order corrections---cf. Eqs.~\eqref{eqn:barym} below---such 
that the power counting in $1/m_{c,b}$ and order-by-order matching of the $F^T_i$ and $G^T_i$ form factors onto HQET requires careful book-keeping.
By contrast, while in deriving the kernel~\eqref{eqn:Kexp} we constructed the amplitude with respect to $q = p - p'$,
the kernel itself is independent of the hadron masses. 
It depends only on $w$, and does not introduce hadron mass terms. 
(Notably, one could have used any timelike momentum to derive the kernel, so long as it was a linear combination of $v \propto p$ and $v' \propto p'$.)
With this in mind, we proceed to show how to match onto HQET using the tensor current definition~\eqref{eqn:HQETffdefTnT},
that manifestly preserves HQET symmetries, while being careful to keep track of the kernel.  

\section{HQET Matching for $\Lambda_b\to\Lambda_c^*$}
\label{sec:match}
\subsection{Spectroscopy}
The $\LcSp$ and $\LcTn$ form a HQS doublet `superfield'~\cite{Leibovich:1997az} 
\begin{equation}
	\psi^\mu_c = \Psic^\mu + \frac{1}{\sqrt{3}}\big(\g^\mu + v^{\prime \mu}\big)\g^5 u_c
\end{equation}
whose `brown muck' degrees of freedom are in the spin-parity $s_\ell^P = 1^-$ state,
while the $\Lb$ has $s_\ell^P = 0^+$ and is an HQS singlet.
The baryon masses can be expressed in HQET via
\begin{align}
	m_{\Lambda_Q} & = m_Q + \LamL - \frac{\lambda_{1,0}}{2m_Q}+ \ldots\,, \label{eqn:barym} \\
	m_{\LSp{Q}} & = m_Q + \LamLp - \frac{\lambda_{1,1}}{2m_Q}  -2\frac{\lambda_{2,1}(m_Q)}{m_Q} + \ldots\,,\nn\\
	m_{\LTn{Q}} & = m_Q + \LamLp - \frac{\lambda_{1,1}}{2m_Q} + \frac{\lambda_{2,1}(m_Q)}{m_Q}+ \ldots\,. \nn
\end{align}
where $Q = c$, $b$, and the ellipsis denotes higher-order terms in $\lqcd/m_Q$.
The parameter $\LamL$ ($\LamLp$) is the energy of the light degrees of freedom 
in the $m_Q\to\infty$ limit for the baryon $s_\ell^P = 0^+$ HQ singlet ($s_\ell^P = 1^-$ HQ doublet), 
and arises also in the semileptonic form factors~\cite{Leibovich:1997tu, Leibovich:1997em}.
The parameters $\lambda_1$ and $\lambda_2$ are related to the HQ kinetic energy and chromomagnetic energy. 

The six baryon masses---two $m_{\Lambda_{c,b}}$ and four $m_{\Lambda^*_{c,b}}$---are well-measured, 
such that when combined with a quark mass scheme for $m_{c,b}$, the remaining six parameters in Eqs.~\eqref{eqn:barym} are fully determined.
As in Refs~\cite{Ligeti:2014kia, Bernlochner:2017jka, Bernlochner:2017jxt,Bernlochner:2018bfn,Bernlochner:2018kxh}, 
we use the $1S$ short-distance mass scheme~\cite{Hoang:1998ng, Hoang:1998hm, Hoang:1999ye},
under which ambiguities in the pole mass, $\LamL$ and $\LamLp$ cancel, and the behavior of the perturbation series is improved. 
Under this scheme, $m_b(m_b^{1S}) \simeq m_b^{1S}(1 + 2\aS^2/9)$, with $m_b^{1S} = (4.71 \pm 0.05)$\,GeV
and we match HQET onto QCD at scale $\mu = \sqrt{m_cm_b}$, so that $\aS \simeq 0.26$. 
Further, the mass splitting $\delta m_{bc} = m_b-m_c = (3.40 \pm 0.02)$\,GeV is well-constrained by $B\to X_c\ell\bar\nu$ spectra~\cite{Bauer:2004ve, Bauer:2002sh},
and is treated as an independent input.
One finds from Eqs.~\eqref{eqn:barym} in particular, $\LamL = 0.81 \pm 0.05$\,GeV and $\LamLp = 1.10 \pm 0.05$\,GeV.
Also important are the HQ expansion parameters $\varepsilon_{c,b} = 1/(2m_{c,b})$.
We summarize in Table~\ref{tab:params} all these inputs and the resulting HQET parameters.

\newcommand{\tcite}[1]{~\text{\cite{#1}}}
\begin{table}[t]
\renewcommand*{\arraystretch}{1.5}
\newcolumntype{D}{ >{\centering\arraybackslash $} m{2.25cm} <{$}}
\newcolumntype{C}{ >{\centering\arraybackslash $} m{1.7cm} <{$}}
\begin{tabular}{CDCD}
	\hline\hline
	\multicolumn{2}{c}{\makecell{\textbf{Inputs} \\ (Masses in GeV)}} & \multicolumn{2}{c}{\makecell{\textbf{HQET Parameters} \\ (Derived)}} \\
	\hline
	m_{\Lb}		& 5.61960(17) \tcite{Zyla:2020zbs}	& \LamL & 0.81(5)\,\GeV \\
	m_{\LSp{b}}  	& 5.91220(12) \tcite{Zyla:2020zbs}	& \LamLp & 1.10(5)\,\GeV\\
	m_{\LTn{b}}  	& 5.91992(19) \tcite{Zyla:2020zbs}	& \lambda_{1,0} & -0.26(7)\,\GeV^2\\
	m_{\Lc}		& 2.28646(14) \tcite{Zyla:2020zbs}	& \lambda_{1,1} & -0.38(7)\,\GeV^2\\
	m_{\LSp{c}}  	& 2.59225(28) \tcite{Zyla:2020zbs}	& \lambda_{2,1}(m_b) & 0.0123(4)\,\GeV^2 \\
	m_{\LTn{c}}  	& 2.62811(19) \tcite{Zyla:2020zbs}	& \lambda_{2,1}(m_c) & 0.0165(7)\,\GeV^2 \\
	m_b^{1S}		& 4.71(5) \tcite{Hoang:1998ng, Hoang:1998hm, Hoang:1999ye}	& \eb[=\frac{1}{2m_b}]& 0.105(1)\,\GeV^{-1} \\
	\delta m_{bc}	& 3.40(2) \tcite{Bauer:2004ve, Bauer:2002sh} & \ec[=\frac{1}{2m_c}] & 0.36(1)\,\GeV^{-1}\\
	\aS(\sqrt{m_bm_c})	& 0.26 	& & \\
	\hline\hline
\end{tabular}
\caption{Input parameters used (left) and derived HQET parameters (right).}
\label{tab:params}
\end{table}

\subsection{Heavy quark expansion}
At leading order in HQET~\cite{Georgi:1990um, Eichten:1989zv}, the matrix elements
\begin{equation}
	\langle \Lcs | \cbar \Gamma b | \Lb \rangle = \sigma(w) v_\alpha \overline{\psi}_c^{\,\alpha} \Gamma u_b\,,
\end{equation}
in which $\sigma$ is the leading order Isgur-Wise (IW) function. 
As for the definition of the form factors, we use (close analogs of) the notation of Ref.~\cite{Leibovich:1997az} for the IW functions.
In the heavy quark limit, matching onto Eqs.~\eqref{eqn:HQETffdefSp}, \eqref{eqn:HQETffdefTn} and~\eqref{eqn:HQETffdefTnT}, 
\begin{align}
	\dP & = \dV1 = \sigma (w-1)/\sqrt{3} \nn \\
	\dS & = \dA1 = \dT1 = \sigma (w+1)/\sqrt{3} \nn \\
	\dV2 & = \dA2 = \dT2 = -2\sigma/\sqrt{3}\,, \nn \\
	\lS & = \lP = \lV1 = \lA1 = \lT1 + \lambda_T = \sigma\,,
\end{align}
and all others vanish. 
Note we have included the $\LcTn$ tensor current kernel parameter $\lambda_T$ as in Eq.~\eqref{eqn:lTredef}. 
One sees here immediately that although $\lambda_T$ is unphysical with respect to the $\Lb \to \LcTn$ amplitudes and is therefore unconstrained, 
the choice $\lambda_T = -\lT1$ is inconsistent with HQET.
That is, one could not have first removed the $\lT1$ term from Eq.~\eqref{eqn:HQETffdefTnT} using the kernel, and then attempted to match onto HQET.

At $\mathcal{O}(1/m_{c,b})$, matching the $\cbar\, \Gamma\, b$ current onto HQET generates current corrections to the matrix element
\begin{equation}
	\langle \Lcs | \delta (\cbar \Gamma b) | \Lb \rangle = -\ec b^{(c)}_{\alpha \mu} \overline{\psi}_c^{\,\alpha} \gamma^\mu \Gamma u_b +\eb b^{(b)}_{\alpha \mu} \overline{\psi}_c^{\,\alpha} \Gamma  \gamma^\mu u_b\,.
\end{equation}
Applying the equations of motion for HQET fields and the heavy quarks, one may show~\cite{Leibovich:1997az}
\begin{align}
	b^{(c)}_{\alpha\mu} & = \sigma_1\big[v_\alpha v_\mu - w v_\alpha v'_\mu + (w^2-1)g_{\alpha\mu}\big] \nn \\
		& \quad + \sigma (\LamL - w\LamLp) g_{\alpha\mu}\,,\nn\\
	b^{(b)}_{\alpha\mu} & = \sigma_1\big[w v_\alpha v'_\mu - v_\alpha v_\mu - (w^2-1)g_{\alpha\mu}\big] \nn \\
		& \quad + \sigma \big[\LamL v_\alpha v_\mu  - \LamLp v_\alpha v'_\mu -  (\LamL - w\LamLp)g_{\alpha\mu}\big]\,,
\end{align}
where $\sigma_1 = \sigma_1(w)$ is a subleading IW function. 
Two additional subleading IW functions $\phi_{c,b}$ arise via $\mathcal{O}(1/m_{c,b})$ chromomagnetic corrections to the HQET Lagrangian,
\begin{multline}
	\delta \mathcal{L} = \ec \pc g_{\alpha\mu} v_\nu \, \overline{\psi}_c^{\,\alpha} i \sigma^{\mn} \frac{1 + \slashed{v}'}{2}\Gamma u_b \\ + 
	\eb \pb g_{\alpha\mu} v'_\nu \, \overline{\psi}_c^{\,\alpha}  \Gamma \frac{1 + \slashed{v}}{2}i \sigma^{\mn}u_b\,.
\end{multline}	
Finally, the kinetic energy operator in the $\mathcal{O}(1/m_{c,b})$ HQET Lagrangian generates HQ spin-symmetry conserving terms
$\big[\LamL \eb \sigma^b_{\text{ke}}(w) + \LamLp\ec \sigma^c_{\text{ke}}(w)\big] v_\alpha \overline{\psi}_c^{\,\alpha} \Gamma u_b$. 
These terms can be consistently reabsorbed into the leading order IW function via the redefinition
\begin{equation}
	\sigma(w) + \big[\LamL \eb\sigma^b_{\text{ke}}(w)   + \LamLp\ec \sigma^c_{\text{ke}}(w) \big] \mapsto \sigma(w)\,.
\end{equation}
It is further convenient to define hatted IW functions and form-factors, normalized by $\sigma$,
\begin{equation}\label{eqn:hatHdef}
	\hat x(w) = x(w) \big/ \sigma(w)\,, \qquad x = \big\{\sigma_1, \phi_{c,b}\,, d_X\,, l_X\,\big\}\,.
\end{equation}

Radiative $\mathcal{O}(\aS)$ corrections to the HQ currents can be computed by matching QCD onto HQET~\cite{Falk:1990yz,Falk:1990cz, Neubert:1992qq}. 
Similar to the notation of Ref.~\cite{Manohar:2000dt}, these corrections arise in form factors via the functions $C_{\Gamma_i}$, where $\Gamma_i$ is a form factor label.
The $C_{\Gamma_i}$ are functions of $w$, $z=m_c/m_b$, and $\haS = \aS/\pi$~\cite{Neubert:1992qq};
explicit expressions for $C_{\Gamma_i}$ are in Ref.~\cite{Bernlochner:2017jka}. 

Incorporating all these contributions, at $\mathcal{O}(1/m_{c,b}, \aS)$, the $\Lb \to \LcSp$ hatted form factors
\begin{widetext}
\begin{subequations}
\label{eqn:exphffSp}
\begin{align}
\sqrt{3} \hdS & = w+1+\Cs (w+1) \haS+\ec \big[3 (\LamLp w - \LamL)-2 (w^2-1)\hs -2 (w+1)\hpc\big] \nn\\
	&\qquad +\eb \big[(w-2) \LamL+(2w-1) \LamLp-2 (w^2-1)\hs +2 (w+1) \hpb\big]\,,\\
\sqrt{3} \hdP & = w-1+\Cps (w-1) \haS+\ec \big[3 (\LamLp w - \LamL)-2 (w^2-1)\hs -2 (w-1)\hpc\big] \nn\\
	&\qquad +\eb \big[(2+w) \LamL-(2 w+1) \LamLp+2 (w^2-1)\hs -2 (w-1) \hpb\big]\,,\\
\sqrt{3} \hdV1 & = w-1+\Cv1 (w-1) \haS+\eb (w \LamL-\LamLp)+\ec \big[3 (\LamLp w - \LamL)-2(w^2-1)\hs -2 (w-1) \hpc\big]\,,\\
\sqrt{3} \hdV2 & = -2 - \big[2 \Cv1+\Cv2 (w+1)\big] \haS+4 \ec \hpc-2 \eb \big[\LamL+\LamLp-(w+1)\hs +\hpb\big]\,,\\
\sqrt{3} \hdV3 & = -\Cv3 (1+w) \haS+2 \eb (\LamL+\LamLp-(w+1) \hs -\hpb)\,,\\
\sqrt{3} \hdA1 & = 1+w+\Ca1 (1+w) \haS+\eb (w \LamL-\LamLp)+\ec \big[3 (\LamLp w - \LamL)-2(w^2-1)\hs -2 (w+1) \hpc\big]\,,\\
\sqrt{3} \hdA2 & = -2-\big[2 \Ca1+\Ca2 (w-1)\big] \haS+4 \ec \hpc+2 \eb \big[\LamLp -\LamL-(w-1)\hs +\hpb\big]\,,\\
\sqrt{3} \hdA3 & = -\Ca3 (w-1) \haS-2 \eb \big[\LamL-\LamLp+(w-1) \hs +\hpb\big]\,,\\
\sqrt{3} \hdT1 & = 1+w+\Ct1 (1+w) \haS+\ec \big[3 (\LamLp w - \LamL)-2 (w^2-1)\hs -2 (w+1)\hpc\big] \nn\\
	&\qquad +\eb \big[(2+w) \LamL-(2 w+1) \LamLp+2 (w^2-1)\hs -2 (w+1) \hpb\big]\,,\\
\sqrt{3} \hdT2 & = -2 - \big[2 \Ct1+\Ct2 (w-1)\big] \haS+4 \ec \hpc+2 \eb \big[\LamLp-\LamL-(w-1)\hs +\hpb\big]\,,\\
\sqrt{3} \hdT3 & = -\Ct3 (w-1) \haS+2 \eb \big[\LamLp-\LamL-(w-1) \hs +\hpb\big]\,,\\
\sqrt{3} \hdT4 & = -2 \Ct3 \haS-4 \eb (\LamLp-w \hs +\hpb)\,.
\end{align}
\end{subequations}
\end{widetext}
Similarly, the $\Lb \to \LcTn$ hatted form factors at $\mathcal{O}(\aS,1/m_{c,b})$
\begin{subequations}
\label{eqn:exphffTn}
\begin{align}
\hlS & = 1+\Cs \haS+\ec \big[(w-1) \hs+\hpc\big] \nn\\
	& \qquad +\eb \big[\LamL-\LamLp+(w-1) \hs-\hpb\big]\,,\\
\hlP & = 1+\Cps \haS+\ec \big[(1+w) \hs+\hpc\big] \nn\\
	& \qquad +\eb \big[\LamL+\LamLp-(w+1) \hs+\hpb\big]\,,\\
\hlV1 & = 1+\Cv1 \haS+\ec \big[(1+w) \hs+\hpc\big] \nn\\
	& \qquad +\eb \big[\LamL+\LamLp-(w+1) \hs+\hpb\big]\,,\\
\hlV2 & = \Cv2 \haS-2 \ec \hs\,,\\
\hlV3 & = \Cv3 \haS-2 \eb \big[\LamLp-w \hs+\hpb\big]\,,\\
\hlV4 & = -2 \eb \!\big[\LamL-w \LamLp+(w^2\!-\!1)\hs-(w-1) \hpb\big]\,,\\
\hlA1 & = 1+\Ca1 \haS+\ec \big[(w-1) \hs+\hpc\big] \nn\\
	& \qquad +\eb \big[\LamL-\LamLp+(w-1) \hs-\hpb\big]\,,\\
\hlA2 & = \Ca2 \haS-2 \ec \hs\,,\\
\hlA3 & = \Ca3 \haS+2 \eb \big[\LamLp-w \hs+\hpb\big]\,,\\
\hlA4 & = 2 \eb \big[\LamL-w \LamLp+(w^2-1)\hs-(w+1) \hpb\big]\,,\\
\hlT1 &  = -\hat\lambda_T + 1+\Ct1 \haS+\ec \big[(w-1) \hs+\hpc\big] \nn\\
	& \qquad +\eb \big[\LamL-\LamLp+(w-1) \hs-\hpb\big]\,,\\
\hlT2 & = \Ct2 \haS-2 \ec \hs\,,\\
\hlT3 &  =  \hat\lambda_T + \Ct3 \haS+2 \eb \big[\LamLp-w \hs+\hpb\big]\,,\\
\hlT4 & =  - (w+1)\hat\lambda_T \\
	& \qquad + 2 \eb \big[\LamL-w \LamLp+(w^2-1)\hs-(w+1) \hpb\big]\,,\nn\\
\hlT5 & = \hat\lambda_T\,,\\
\hlT6 &  = \hat\lambda_T + 4 \eb \hpb\,,\\
\hlT7 & = 0\,,
\end{align}
\end{subequations}
in which we have kept the tensor current kernel terms to the HQET expansion, with $\hat\lambda_T = \lambda_T/\sigma$.

The results for the vector and axial-vector form factors match those of Ref.~\cite{Leibovich:1997az}, 
while the (pseudo)scalar and tensor form factors are new.
We see in particular that if one had already fixed $\lambda_T = \lT3$, $\lambda_T = \lT6$, or $\lambda_T = -\lT4/(w+1)$
in order to eliminate their respective terms from Eq.~\eqref{eqn:HQETffdefTnT}, then this would have been incompatible with matching onto HQET at $\mathcal{O}(\aS,1/m_{c,b})$.
That is, the HQ expansion~\eqref{eqn:exphffTn} generates relations between the form factors, 
thus providing HQET constraints on the choice of $\lambda_T$ in order to eliminate any given tensor current term.
In particular, eliminating the $\lT6$ term~\cite{Mott:2011cx} would require $\lambda_T = -4\eb \pb$.

\section{HQET parametrization fits}
\label{sec:fits}
In Appendix~\ref{app:ampl} we present the full set of NP helicity amplitudes and decay rates for $\Lb \to \Lcs l \nu$.
Just as for the parametrization of the IW functions for $B \to D^{**}$ decays~\cite{Bernlochner:2016bci,Bernlochner:2017jxt},
we parametrize the IW functions to at most linear order in $(w-1)$ as
\begin{align}
	\sigma(w) & \simeq \sigma(1)\big[1 + \sigma'(w-1) \big]\,\nn\\
	 \hs(w) & \simeq \hs(1)\,, \nn\\ 
	 \hpc(w) & \simeq \hpc(1)\,, \label{eqn:HQETparam}
\end{align}
and $\hpb(w) \simeq 0$. (Ref.~\cite{Boer:2018vpx} implicitly sets both $\phi_{c,b} = 0$, because therein chromomagnetic corrections are dropped.)
At present, there is no available experimental data which would permit data-driven fits over the full recoil range to these HQET parameters $\sigma(1)$, $\sigma'$, $\hs(1)$, and $\hpc(1)$.
A measurement of their relative branching fractions to $\Lb \to \Lc l \nu$ is, however, available~\cite{Aaltonen:2008eu} for the ratios
\begin{align}
	f(\Lcs) & \equiv \frac{\Gamma[\Lb \to \Lcs \mu \nu]}{\Gamma[\Lb \to \Lc \mu \nu]}\,,\nn\\
	f\big(\LcSp\big) & = 0.126 \pm 0.033^{+0.047}_{-0.038}\,,\nn\\
	f\big(\LcTn\big) & = 0.210 \pm 0.042^{+0.071}_{-0.050}\,. \label{eqn:RLcsLc}
\end{align}
Furthermore, very recently LQCD results have been published~\cite{Meinel:2021rbm}, applicable only in the near-zero recoil regime $w \le 1.05$.

\subsection{Quark models}

As a preliminary measure, one can fit the parametrization~\eqref{eqn:HQETparam} to quark-model based predictions for the vector and axial-vector currents over the full $w$ range,
with the intent that such fits can provide a plausible set of initial central values for the HQET parameters, for use in future data-driven fits.
To this end, we make use of the results of the PCR form factor parametrization~\cite{Pervin:2005ve} for $\Lb \to \Lcs$,
which is based on a constituent quark model, and follows a similar approach to ISGW2~\cite{Scora:1995ty,Isgur:1988gb}.
For each decay mode, explicit expressions for the (axial-)vector form factors are obtained 
in terms of two wave function overlap parameters $\alpha_{\lambda}$, $\alpha_{\lambda'}$, 
along with effective masses for the heavy quarks, $\bar{m}_{c,b}$, and an effective light quark mass, $\bar{m}_\sigma$: a total of five parameters.
We use the same numerical values as implemented in \texttt{EvtGen R01-07-00}~\cite{Lange:2001uf}, shown in Table~\ref{tab:PCR}.
Using data-driven precision predictions for $\Lb \to \Lc l \nu$~\cite{Bernlochner:2018kxh}, 
this PCR parametrization predicts $f\big(\LcSp\big) \simeq 0.13$ and $f\big(\LcTn\big) \simeq 0.26$ in good agreement with the data~\eqref{eqn:RLcsLc}.

\begin{table}[t]
\renewcommand*{\arraystretch}{1.5}
\newcolumntype{C}{ >{\centering\arraybackslash $} m{1.5cm} <{$}}
\newcolumntype{D}{ >{\centering\arraybackslash $} m{1cm} <{$}}
\begin{tabular}{DDCCC}
	\hline\hline
	\alpha_\lambda & \alpha_{\lambda'} & \bar{m}_b\,[\GeV] & \bar{m}_c\,[\GeV] & \bar{m}_\sigma\,[\GeV] \\
	\hline
	 0.59 & 0.47 & 5.28 & 1.89 & 0.40\\
	\hline\hline
\end{tabular}
\caption{Parametric values for the PCR form-factor parametrization, as implemented in \texttt{EvtGen R01-07-00}~\cite{Lange:2001uf},
for both $\LcSp$ and $\LcTn$.}
\label{tab:PCR}
\end{table}

To fit the HQ expansion of the form factors to the PCR parametrization, 
we sample both of its predicted $d\Gamma/dw$ spectra at three points $w = 1.1$, $1.2$, and~$1.3$.
(The maximum $w_{\text{max}} = 1.315$ and $1.303$ for $\Lb \to \LcSp \ell \nu$ and $\LcTn \ell \nu$, respectively. Hereafter $\ell = e$ or $\mu$.)
We further assign a somewhat arbitrary $20\%$ theory uncertainty to this PCR data, partly informed by the uncertainties in Ref.~\cite{Pervin:2005ve}.
This is quite a bit smaller than the $40$--$45\%$ uncertainties in $f(\Lcs)$ measurements~\eqref{eqn:RLcsLc}, 
so that the latter are not included in the fit as they would generate only relatively weaker pulls.

One obtains from a simultaneous fit of the $\Lb \to \LcSp \ell \nu$ and $\LcTn \ell\nu$ spectra to the six PCR data points
\begin{align}
	\sigma(1) & \sim 1.0\pm 0.1\,,  & \sigma'  \sim -1.8 \pm 0.2\,, \nn\\
	 \hs(1) & \sim 0.9 \pm 0.5\,,	&  \hpc(1)  \sim  0.2 \pm 0.3\,, \label{eqn:PCRfitres}
\end{align}
in which we emphasize here (and hereafter) by the `$\sim$' that these fit results and their uncertainties 
do not properly include the (presumably quite large) theory uncertainties or possible biases implicit to the 
quark model-dependent PCR parametrization predictions and the sampling thereof. 
Moreover we emphasize these results are not data-driven. 
We therefore refer to these results as `fit estimates' for the remainder of this discussion.
Sensitivity to the charm chromomagnetic contribution in this fit is weak, with $\hpc(1)$ compatible with zero, but not entirely negligible.

\begin{figure}[tb]
	\includegraphics[width = 8cm]{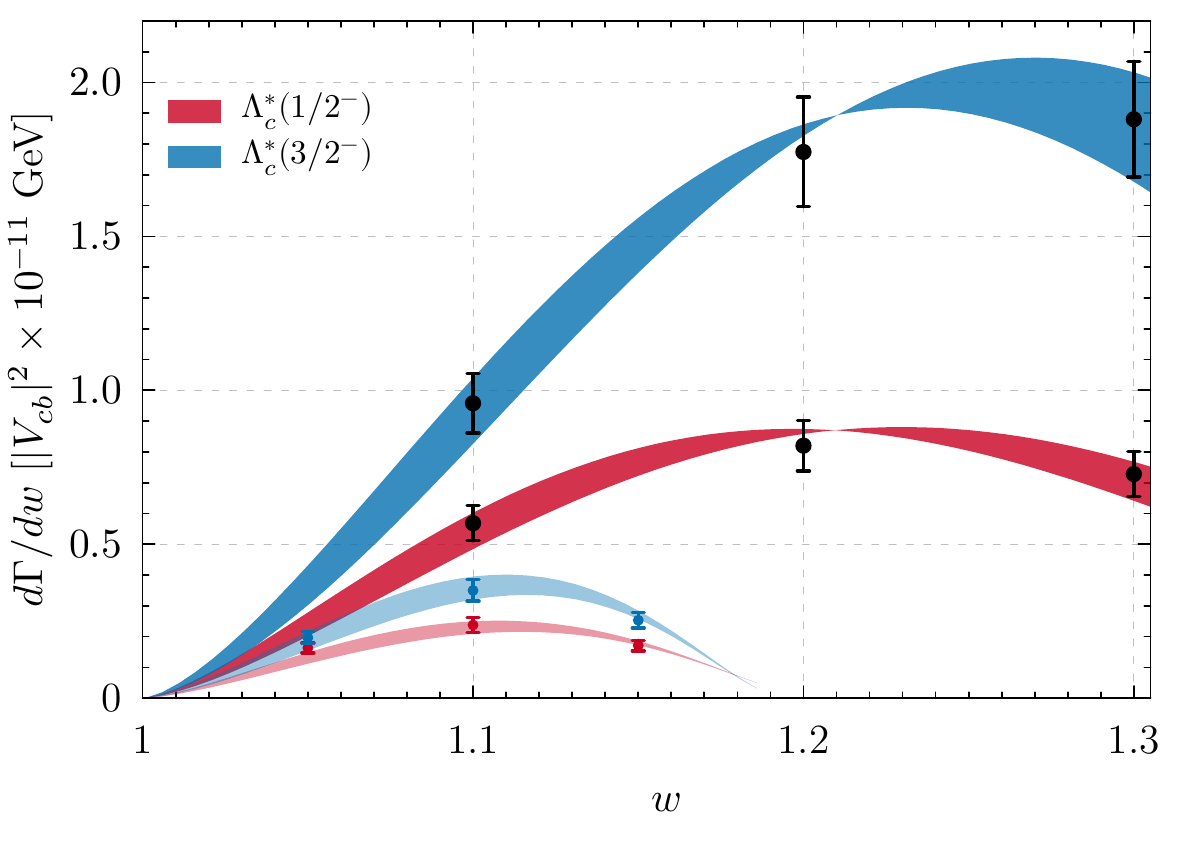}
	\caption{$d\Gamma/dw$ spectra for $\Lb \to \LcSp \ell \nu$ (red band) and $\LcTn \ell \nu$ (blue band) using the fit estimates~\eqref{eqn:PCRfitres}, 
	obtained from fitting the HQET expansion parameters to data points generated by predictions of the PCR parametrization~\cite{Pervin:2005ve} (black points).
	Each spectrum is shown as a band, generated by the first principal component and uncertainty of the fit.
	In lighter corresponding colors we show the respective semitauonic decay distributions, and their PCR predictions at $w = 1.05$, $1.1$ and $1.15$.
	}
	\label{fig:FFPCR}
\end{figure}

In Fig.~\ref{fig:FFPCR} we show the resulting estimated $w$ spectra for $\Lb \to \LcSp \ell\nu$ and $\LcTn \ell \nu$, compared to the predicted PCR data points.
The spectra are displayed as red and blue bands, respectively, that show the coverage by the first principal component and uncertainty of the fit covariance.
The fit estimates predict the ratios $f(\LcSp) \sim 0.130(3)$ and $f(\LcTn) \sim 0.259(4)$ in good agreement with the data~\eqref{eqn:RLcsLc}.
In lighter red and blue bands we show the corresponding $\Lb \to \LcSp \tau\nu$ and $\LcTn \tau \nu$ spectra, respectively. 
These are in good agreement with PCR predictions, shown as respectively colored red and blue data points sampled at $w= 1.05$, $1.1$ and $1.15$.
If instead one uses these six $\Lb \to \LcSp \tau\nu$ and $\LcTn \tau \nu$ data points as additional fit points 
together with the six $\Lb \to \LcSp \ell\nu$ and $\LcTn \ell \nu$ data points, 
one finds $\sigma(1) \sim 0.99 \pm 0.04$, $\sigma'  \sim -1.8 \pm 0.1$, $\hs(1) \sim 1.0 \pm 0.3$, and $ \hpc(1)  \sim  0.1 \pm 0.2$, 
which is entirely compatible with the fit estimates in Eq.~\eqref{eqn:PCRfitres}.

\subsection{Comparison with LQCD predictions}\label{sec:LQCD}
We now turn to examine the recent LQCD results~\cite{Meinel:2021rbm}, and HQET fits thereto. 
First, in the HQ limit the ratio of the differential decay rates to the $\Lcs$ HQ doublet states
\begin{equation}
	\frac{d\Gamma[\Lb \to \LcSp l \nu]/dw}{d\Gamma[\Lb \to \LcTn l \nu]/dw} = \frac{1}{2}\,,
\end{equation}
based on the number of spin degrees of freedom in the respective final states, and irrespective of any NP present.
This is in good agreement with the ratio of the measured branching fractions, which we compute from Eq.~\eqref{eqn:RLcsLc} to be 
\begin{equation}
	\label{eqn:RfLcs}
	R_f  \equiv \frac{\Gamma[\Lb \to \LcSp \mu \nu]}{\Gamma[\Lb \to \LcTn \mu \nu]} = \frac{f(\LcSp)}{f(\LcTn)} = 0.6^{+0.4}_{-0.2}\,. 
\end{equation}
This, in turn, indicates the absence of large HQS violations, at least after integrating over the full phase space.

However, it is notable that the LQCD results instead predict $d\Gamma[\Lb \to \LcSp l \nu]/dw \sim 2.5\times d\Gamma[\Lb \to \LcTn l \nu]/dw$, albeit for $w \le 1.05$ only.
That is, in the near-zero recoil region the ratio of the $\LcSp$ mode to the $\LcTn$ mode is five times larger than would be expected from the HQ limit.
This alone suggests large HQS breaking in the LQCD results,
naively far larger than the expected effects from potentially large $1/m_{c}^2$ terms. 
This is, perhaps, a surprising result given that the HQ expansion to $\mathcal{O}(1/m_c^2)$, when used for the description of the ground state $\Lb \to \Lc \ell \nu$ decays,
shows excellent agreement with SM LQCD results and experimental data~\cite{Bernlochner:2018kxh}.
(A moderate discrepancy, however, between the LQCD prediction for the $\Lb \to \Lc$ tensor form factors, 
and the predictions from a combined fit of the $\Lb \to \Lc \ell \nu$ experimental data and SM LQCD results, 
has been noted previously~\cite{Bernlochner:2018bfn}.)

The LQCD results are expressed in terms of a particular form factor helicity basis for the vector, axial vector, tensor and axial tensor form factors, 
in which each is parametrized to linear order as $F_f + A_f(w-1)$, i.e. two parameters for each form factor.
The transformation from the HQET basis~\eqref{eqn:HQETffdefSp},~\eqref{eqn:HQETffdefTn} and~\eqref{eqn:HQETffdefTnT} 
to this LQCD helicity basis is provided in Appendix~\ref{app:FFbasis}, 
with the explicit definitions in Ref.~\cite{Meinel:2021rbm} (see also e.g. Refs.~\cite{Descotes-Genon:2019dbw,Meinel:2020owd}).
Results for the scalar and pseudoscalar form factors are not explicitly provided; there are therefore a total of $(10 + 14)\times 2 = 48$ parameters in the LQCD results.
For this discussion, we use the nominal fit parameters in Table VII of Ref.~\cite{Meinel:2021rbm}, and covariance matrices provided therein.

We match HQET onto QCD at $\mu = \sqrt{m_c\,m_b}$, while the LQCD results are taken to be computed at $\mu = m_b$ (cf. Ref.~\cite{Datta:2017aue}).
To account for the anomalous dimension of the tensor current, 
we include a multiplicative renormalization factor $[\alpha_s(m_cm_b)/\alpha_s(m_b)]^{-4/25} \simeq 0.97$~\cite{Dorsner:2013tla,Freytsis:2015qca}
for each LQCD tensor form factor, in order to scale them to our matching scale.
The 24 LQCD form factors are each sampled at two $w$ values, $w=1$ and $w=1.05$, 
in order to create a set of 48 correlated, nonredundant data points over the joint space of $w$ and the form factors.
Fitting the five-parameter HQET parametrization~\eqref{eqn:HQETparam}	 of the form factors---i.e. including $\hpb(w)$---to 
these data, we find a dramatically poor fit, with $\chi^2/\text{dof} \simeq 313/43$.
Turning off $1/m_{c,b}$ and $\alpha_s$ corrections worsens the $\chi^2$ of the fit, as expected, but only marginally; 
not accounting for the $\alpha_s$ running described above significantly worsens the fit, with $\chi^2/\text{dof} \simeq 484/43$.

To roughly characterize the degree of HQS breaking implied by the LQCD results, 
we modify the fit to include HQS-breaking parameters for each hatted HQET basis form factor 
in Eqs.~\eqref{eqn:exphffSp} and~\eqref{eqn:exphffTn}:
i.e. $\hat{d}_X \to \hat{d}_X + \varepsilon_{d_X}$ and $\hat{l}_X \to \hat{l}_X + \varepsilon_{l_X}$. 
In this $5 + 24$ parameter fit, an excellent fit to the LQCD results is obtained, with $\chi^2/\text{dof} \simeq 25.3/19$.
However, some of these HQS-breaking parameters are pulled to very large values,
e.g. we find $\varepsilon_{\dV2} = 0.7 \pm 0.2$ and $\varepsilon_{\dA2} = 0.8 \pm 0.2$: 
The best fit values and the uncertainties of the HQS-breaking parameters are reported in Appendix~\ref{app:nuisance}.
Since by comparison $[\LamLp \varepsilon_c]^2 \simeq 0.16$ and $[\LamL \varepsilon_c]^2 \simeq 0.085$, 
this also suggests the presence of very large HQS breaking effects in the LQCD results,
perhaps beyond what could be expected from large $1/m_c^2$ terms alone.
Interestingly, the $\LcTn$ HQS-breaking parameters are all smaller, being either compatible with zero or the naively-expected size of $1/m_c^2$ terms.

Using the modified fit results with the HQS-breaking parameters and naively extrapolating over the whole $w$ range, 
one can compute the ratio of branching fractions, $R_f$, between the $\LcSp$ and $\LcTn$ modes,
obtaining $R_f = 3.4 \pm 1.1$. 
This is in tension with the ratio of the experimental measurements in Eq.~\eqref{eqn:RfLcs}.
If these results are confirmed, the only way to reconcile the LQCD results with the experimental data
would be non-trivial $w$-dependence of the HQS-violating terms, 
such that very large HQS violations in the $w \lesssim 1.05$ region are balanced out elsewhere in the $w$ spectrum. 
To further investigate this, a full fit to the LQCD results incorporating higher-order $\mathcal{O}(1/m_c^2, \aS/m_{c,b})$ 
corrections in HQET is warranted, though beyond the scope of the current work.

\subsection{SM and NP predictions}

Returning to the quark model fit estimates~\eqref{eqn:PCRfitres}, the corresponding lepton universality ratios, defined by 
\begin{equation}
	R(\Lcs) = \frac{\Gamma[\Lb \to \Lcs \tau \nu]}{\Gamma[\Lb \to \Lcs \ell \nu]}\,, \qquad \ell = e\,,\mu\,,
\end{equation}
are predicted to be in the SM
\begin{equation}
	R\big(\LcSp\big) \sim 0.15\pm0.01\,,\quad R\big(\LcTn\big) \sim 0.11 \pm 0.01\,,
\end{equation}
in which again we emphasize by the `$\sim$' that the estimates feature unassessed theory uncertainties, and are not data-driven.
This can be compared to the PCR SM predictions $R(\LcSp) \simeq 0.16$ and $R(\LcTn) \simeq 0.11$.
The quark-model based predictions of Refs~\cite{Gutsche:2018nks,Becirevic:2020nmb} are also compatible with these estimates.

\begin{figure}[t]
\includegraphics[width=8cm]{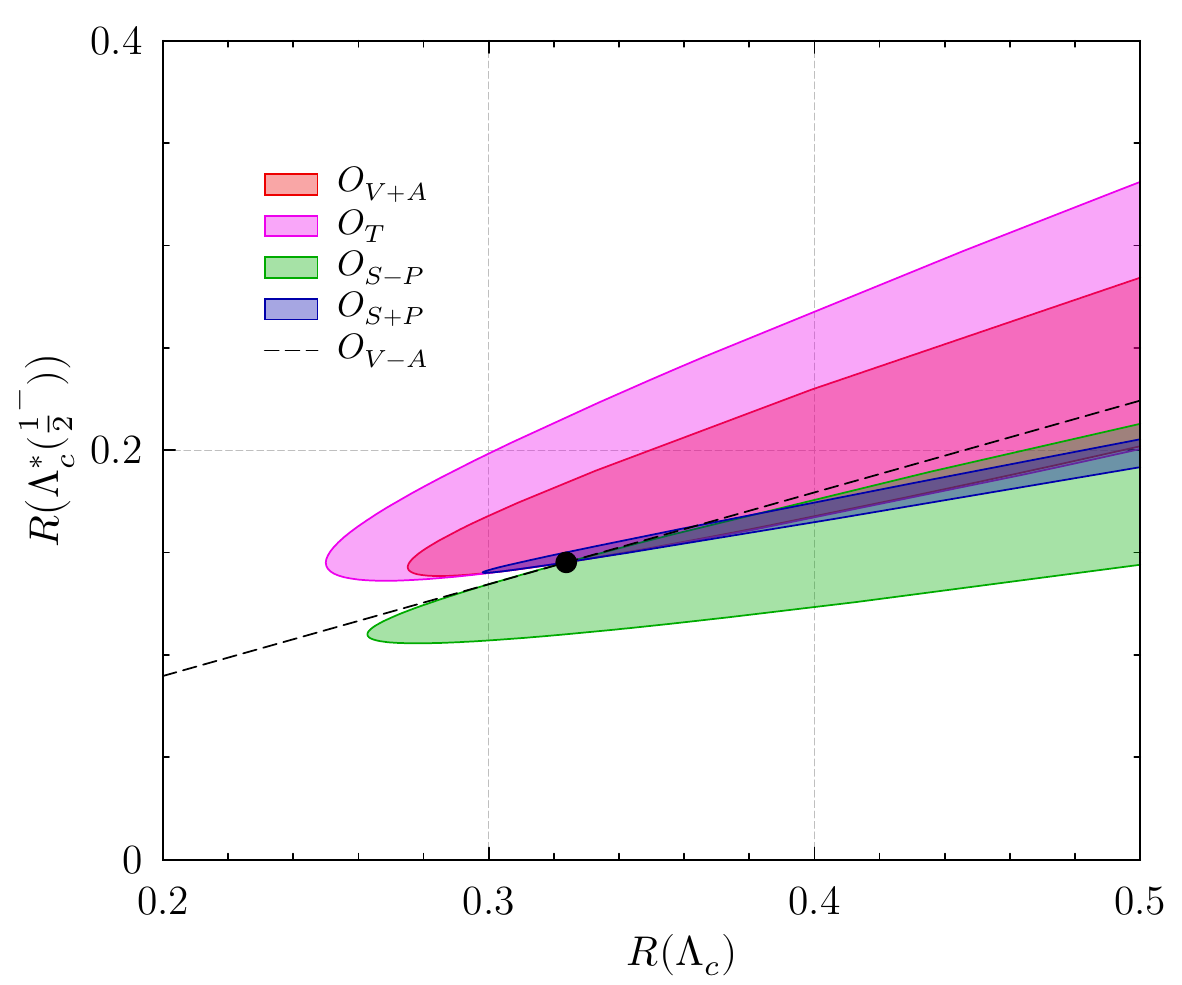}
\includegraphics[width=8cm]{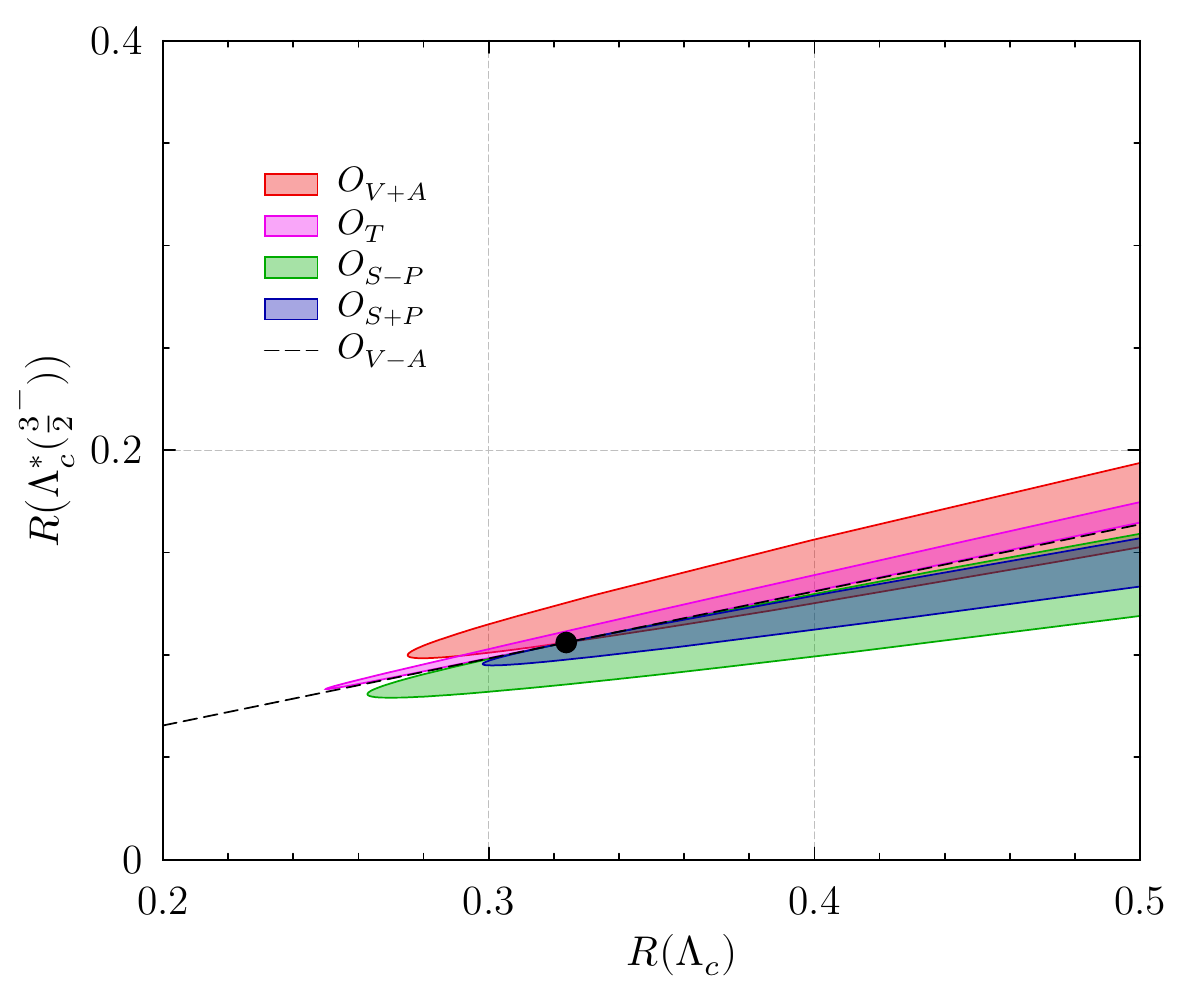}
\caption{Allowed regions in the $R(\Lc)$--$R(\LcSp)$ plane (top) and $R(\Lc)$--$R(\LcTn)$ plane (bottom) for various NP operators combinations.
The SM prediction is shown by a black dot.}
\label{fig:NPRs}
\end{figure}

Using the HQET expansion for the NP form factors, we can immediately extend these fit estimates~\eqref{eqn:PCRfitres} to generate NP predictions for $R(\Lcs)$. 
In Fig.~\ref{fig:NPRs}, we show the allowed regions in the $R(\Lc)$--$R(\Lcs)$ plane for both the $\LcSp$ and $\LcTn$, 
as any one of the NP couplings for the five operator combinations 
$\mathcal{O}_{S+P}$, $\mathcal{O}_{S-P}$, $\mathcal{O}_{V+A}$, $\mathcal{O}_{V-A}$ and $\mathcal{O}_T$ are turned on, 
and assuming only left-handed SM neutrinos. 
These operators enter the effective $b \to c \tau \nu$ Lagrangian as
\begin{equation}
	\label{eqn:Leff}
	\mathcal{O}_X  \sim (\bar{c} \Gamma_X b) (\bar\tau \Gamma_X' \nu)\,,
\end{equation}
with respectively $\Gamma_X = P_R$, $P_L$, $\gamma^\mu P_R$, $\gamma^\mu P_L$ and $\sigma^{\mu\nu}P_L$, 
and $\Gamma_X' = P_L$, $P_L$, $\gamma_\mu P_L$, $\gamma_\mu P_L$ and $\sigma_{\mu\nu} P_L$.
The NP predictions for $R(\Lc)$ are from Ref.~\cite{Bernlochner:2018bfn}.
The boundary of each region corresponds to the case that the NP Wilson coefficients are relatively real with respect to the SM term, 
while the interior requires a relative phase between the SM and NP contributions.  
The $V-A$ NP interaction simply rescales the SM current, and therefore spans a line in the $R(\Lc)$--$R(\Lcs)$ plane.
It is notable that the allowed regions for $\LcSp$ ($\LcTn$) appear moderately (tightly) positively correlated with $R(\Lc)$.

\section{Summary}

We have computed $\Lb \to \Lcs l\nu$ amplitudes for general NP contributions, 
and the corresponding form factors to $\mathcal{O}(1/m_{b,c},\alpha_s)$ in HQET, to be included in the \hammer library~\cite{Bernlochner:2020tfi,bernlochner_florian_urs_2020_3993770}. 
This will provide experimental collaborations with the necessary tools to control the uncertainties associated with the $\Lb \to \Lcs$ feed-down backgrounds in $\Lb \to \Lc l \nu$ analyses.
Moreover, such tools also open the possibility of using the $\Lb \to \Lcs$ modes to directly probe NP or measure $|V_{cb}|$.

Since there are no published differential measurements of $\Lambda_b\to\Lcs l\nu$ decays, 
we have fitted a parameterization of the relevant Isgur-Wise functions to the ISGW2-like PCR parameterization~\cite{Pervin:2005ve}. 
The resulting preliminary fit estimates might be plausibly used as central values for a future data-driven fit,
and we have used them to generate preliminary predictions for the lepton universality ratios $R(\Lcs)$ in both the SM and beyond.
Further, we have assessed the compatibility of the HQET form factors with total decay rates measurements and recent LQCD predictions~\cite{Meinel:2021rbm} in the near-zero recoil regime: 
As previously noted~\cite{Meinel:2021rbm}, there is a tension between LQCD data and HQET predictions. 
We find that this tension persists at full $\mathcal{O}(1/m_{b,c},\alpha_s)$, 
and we furthermore find the LQCD results indicate unexpectedly large HQS-violating terms---potentially, large $1/m_c^2$ corrections---near zero recoil.
However, such large HQS-violating terms cannot persist uniformly over the full recoil spectrum as they would be incompatible with the measurement of the ratio of total decay rates to the $\LcSp$ and $\LcTn$ states. 
Further studies of this issue are warranted and we leave them to the future.

From a more field-theoretical perspective, we saw that the $\Lb \to \Lcs$ system offers an (to our knowledge, first) example 
where the most general parameterization of the hadronic matrix elements compatible with all the symmetries of HQET is overcomplete 
with respect to the number of physical amplitudes: 
There appear to be more form factors than partial wave amplitudes. 
Specifically, this happens in the (pseudo)tensor matrix element for the $\Lb \to \LcTn$ transition. 
Preserving such redundancy allows one to perform manifestly consistent calculations while keeping the HQET symmetries manifest order-by-order in the HQ expansion. 
Therefore, we have presented a general discussion of the counting of the physical form factors for any hadronic transition matrix element,
showing how to determine the physical basis of form factors and match them onto HQET.

\medskip
\textbf{Note added in the proofs:} 
The CDF data reported in Eq.~\eqref{eqn:RLcsLc} assume isospin limit relations for $\LcSp \to \Lambda \pi^+ \pi^-$ 
versus $\LcSp \to \Lambda \pi^0 \pi^0$, using measurements only of the former to recover the $f(\LcSp)$ ratio. 
Significant isospin violation from thresholds in $\LcSp \to \Sigma_c(2455) \pi \to \Lambda \pi \pi$ may alter the recovered ratio, 
but with theory uncertainties that are not yet understood. 
See Ref.~\cite{Zyla:2020zbs} and references therein.
We thank Marcello Rotondo for pointing this out.

\acknowledgements{
We thank Patrick Owen and Mark Wise for helpful discussions, 
and Florian Bernlochner, Marat Freytsis, and Zoltan Ligeti for discussions and comments on the manuscript. 
MP is supported by the Walter Burke Institute for Theoretical Physics.
DJR is supported in part by the Office of High Energy Physics of the U.S. Department of Energy under contract DE-AC02-05CH11231.
}

\appendix
\section{Amplitudes}
\label{app:ampl}
We write explicit expressions for the $\bbar \to \cbar$ amplitudes rather than $b \to c$, defining the basis of NP operators to be
\begin{subequations}\label{abdef}
\begin{align}
\text{SM:\,} & \phantom{-}2\sqrt{2}\, V_{cb}^* G_F\big[\bbar \g^\mu P_L c\big] \big[\bar\nu \g_\mu P_L l\big]\,, \\*
\text{Vector:\,} & \phantom{-}2\sqrt{2}\, V_{cb}^* G_F
  \big[\bbar\big(\alVL \g^\mu P_L + \alVR \g^\mu P_R\big)c\big] \nn\\
  & \quad \times \big[\bar\nu\big(\beVL \g_\mu P_L + \beVR \g_\mu P_R\big) l\big]\,, \\
\text{Scalar:\,} & -2\sqrt{2}\, V_{cb}^* G_F
  \big[\bbar\big(\alSL P_L + \alSR P_R\big)c\big] \nn\\
  & \quad \times \big[\bar\nu\big(\beSL P_R + \beSR P_L\big) l\big]\,, \\
\text{Tensor:\,} & -2\sqrt{2}\, V_{cb}^* G_F
  \big[ \big(\bbar \alTR \sigma^\mn P_R c\big)
  \big(\bar\nu \beTL \sigma_\mn P_R l \big)   \nn\\
  & \quad + \big(\bbar \alTL \sigma^\mn P_L c\big)
  \big(\bar\nu \beTR \sigma_\mn P_L l \big) \big]\,.
\end{align}
\end{subequations}
The subscript of $\beta$ denotes the $\nu$ chirality and the subscript of $\alpha$ is that of the $c$ quark.
Operators for the CP conjugate $b \to c$ processes follow by Hermitian
conjugation.
(The correspondence between the $\alpha$, $\beta$ coefficients and the basis typically chosen for $b\to c$ operators can be found in Ref.~\cite{Bernlochner:2017jxt}.)
The $\Lambda_b \to \Lambda_c^* l \nu$ process has four external
spins: $s_b=\pm$, $s_l=1,2$, $s_\nu=\pm$ and $s_c=1,2$ or $s_c = \pm\frac{3}{2}$, $\pm\frac{1}{2}$ for the spin-$1/2$ or spin-$3/2$ states, respectively.   
(Anticipating inclusion of $\LcSp$ and $\tau$ decays, we label the $\LcSp$ and massive lepton spin by `$1$' and `$2$', 
rather than `$-$' and `$+$', to match the conventions of Ref.~\cite{Ligeti:2016npd} for massive spinors on internal lines.)

\begin{figure}[tb]
	\includegraphics[width = 0.4\linewidth]{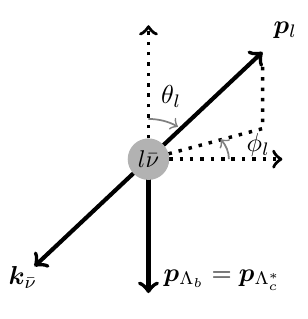}
	\caption{Definition of the helicity angles $\thtau$ and $\phtau$ in the lepton system rest frame. 
	The azimuthal angle $\phtau$ is unphysical in the pure $\Lb \to \Lcs l \nu$ decay.}
	\label{fig:polardef}
\end{figure}

Helicity angles are defined with respect to the $\bbar \to \cbar$ process;
definitions for the conjugate process follow simply by replacing all particles with their antiparticles. 
The single physical polar helicity angle, $\thtau$, defines the orientation of $\bm{p}_l$ in the lepton center of mass reference frame, 
with respect to $-\bm{p}_{\Lambda_b}$, as shown in Fig.~\ref{fig:polardef}.
If subsequent $\Lambda_c^* \to Y_1 \ldots Y_n$ decays are included, one may further define $\phtau$ and $\phi_{i (i+1)}$, as twist angles of the
$l$--$\nu$ and $Y_i$--$Y_{i+1}$ decay planes, such that the combination $\phtau - \phi_{i(i+1)}$ becomes a physical phase. 
For example, including the $\Lambda_c^* \to \Lambda_c \pi \pi$ decay would result in two physical phases. 
Our phase conventions match the spinor conventions of Ref.~\cite{Ligeti:2016npd} for not only $\tau$ but also $\Lambda_c^*$ decay amplitudes,
and are chosen such that the $\Lambda_b \to \Lambda_c^* l \nu$ amplitudes themselves are independent of $\phtau$.

This phase convention amounts to requiring the inclusion an additional spinor phase function in the $\tau$, $\LcSp$ and $\LcTn$ decay amplitudes,
that modifies the phase conventions of the usual helicity basis for the $\tau$ and $\Lcs$.
These phase functions are denoted $h^l_{s_l s_\nu}$, $h^{1/2}_{s_c s_b}$ and $h^{3/2}_{s_c s_b}$, respectively, 
each obeying $h_{\bar{s}\bar{r}} = h_{s r}^*$.
Explicitly the required additional phase factors are 
\begin{align}
	h^l_{1 \dn}(\phtau) & = 1\,, &  h^l_{1 \up}(\phtau) & = e^{i\phtau}\,, \label{eqn:hdefs}\\
	h^{1/2}_{1 \dn}(\phtau) & = 1\,, &  h^{1/2}_{1 \up}(\phtau) & = e^{-i\phtau}\,, \nn\\
	h^{3/2}_{-3/2\, \dn}(\phtau) & = e^{2i\phtau}\,, &  h^{3/2}_{-1/2\, \dn}(\phtau) & = e^{i\phtau}\,, \nn\\
	h^{3/2}_{1/2 \dn}(\phtau) & = 1\,, & h^{3/2}_{3/2 \dn}(\phtau) & = e^{-i\phtau}\,,\nn
\end{align}
plus conjugates.

For compact expression of the amplitudes, it is convenient to define 
\begin{equation}
	w_\pm = w \pm \sqrt{w^2-1}\,, \qquad \mSqq = q^2/m_{\Lb}^2 = 1 - 2 \rC w + \rC^2\,,
\end{equation}	
with $\rC = m_{\Lcs}/m_{\Lb}$, $\rt = m_l/m_{\Lb}$ and further
\begin{align}
\Sigma_{\pm} & = \sqrt{\Wp} \pm \sqrt{\Wm}\,, \nn \\
 R_{\pm} &= (\rC \pm 1)\Sigma_{\mp} + (\rC \mp 1) \Sigma_{\pm}\cos\thtau\,,\nn\\
\Omega_{+} &= \rC - w + \sqrt{w^2 -1}\, \cos\thtau\,, \nn \\ 
\Omega_{\times} &= \rC w- 1 + \rC\sqrt{w^2 -1}\, \cos\thtau\,, \nn \\
 \Omega_{0} & = \sqrt{w^2 -1} + (\rC - w)\, \cos\thtau\,.
\end{align}

We pull out a prefactor $2\sqrt{2}G_F m_{\Lambda_b}^2\sqrt{\rC(\mSqq -\rt^2)}$ from the amplitudes, $A_{s_b s_c s_l s_\nu}$, such that the total differential rate
\begin{multline}
	\frac{d^2\Gamma}{d w \, d\cos\thtau} = \frac{G_F^2 m_{\Lb}^5 \rC^3}{32\pi^3}\sqrt{w^2-1}\frac{(\mSqq - \rt^2)^2}{\mSqq} \\ 
		\times \sum_{s_b,s_c,s_l,s_\nu}\!\!\!\Big| A_{s_b s_c s_l s_\nu}\Big|^2\,. \label{eqn:totdiff}
\end{multline}
The explicit amplitudes for $\Lb \to \LcSp l \nu$ are then
\begin{subequations}
\begin{align}
& A_{\dn\,1\,1\,\dn}  = \bigg\{\frac{1}{2} \beSL (\dP (\alSL-\alSR) \WpmP-\dS (\alSL+\alSR) \WpmM) \nn \\
 		\quad & +\frac{\Rp \dV1 (1+(\alVL+\alVR) \beVL) \rt}{2 \mSqq} \nn \\
 		\quad & +\frac{(1+(\alVL+\alVR) \beVL) \rt \WpmM (\dV3 \Op+\dV2 \Ot)}{2 \mSqq} \nn \\
 		\quad & +\frac{\Rm \dA1 (1+(\alVL-\alVR) \beVL) \rt}{2 \mSqq} \nn \\
 		\quad & +\frac{(1+(\alVL-\alVR) \beVL) \rt \WpmP (\dA3 \Op+\dA2 \Ot)}{2 \mSqq} \nn \\
 		\quad & +4 \dT1 \alTR \beTL \sqrt{\Wp} \cos\thtau \nn \\
 		\quad & +2 \alTR \beTL \big[(\dT2+\dT3) \WpmP-\dT4 \sqrt{w^2 - 1} \WpmM\big] \cos\thtau\bigg\}\\
& A_{\dn\,1\,2\,\dn}  = \sin\thtau\bigg\{-\frac{(\rC - 1) \dV1 (1+(\alVL+\alVR) \beVL) \WpmP}{2 \sqrt{\mSqq}} \nn \\
 		\quad & -\frac{(\rC \dV2+\dV3) (1+(\alVL+\alVR) \beVL) \sqrt{w^2 - 1} \WpmM}{2 \sqrt{\mSqq}} \nn \\
 		\quad & -\frac{(1+\rC) \dA1 (1+(\alVL-\alVR) \beVL) \WpmM}{2 \sqrt{\mSqq}} \nn \\
 		\quad & -\frac{(\rC \dA2+\dA3) (1+(\alVL-\alVR) \beVL) \sqrt{w^2 - 1} \WpmP}{2 \sqrt{\mSqq}} \nn \\
 		\quad & -\frac{4 \dT1 \alTR \beTL \rt \sqrt{\Wp}}{\sqrt{\mSqq}} \nn \\
 		\quad & +\frac{2 \alTR \beTL \rt \big[-(\dT2+\dT3) \WpmP+\dT4 \sqrt{w^2 - 1} \WpmM\big]}{\sqrt{\mSqq}}\bigg\}\\
& A_{\dn\,1\,1\,\up}  = \sin\thtau\bigg\{-\frac{(\rC - 1) \dV1 (\alVL+\alVR) \beVR \WpmP}{2 \sqrt{\mSqq}} \nn \\
 		\quad & -\frac{(\rC \dV2+\dV3) (\alVL+\alVR) \beVR \sqrt{w^2 - 1} \WpmM}{2 \sqrt{\mSqq}} \nn \\
 		\quad & -\frac{(1+\rC) \dA1 (\alVL-\alVR) \beVR \WpmM}{2 \sqrt{\mSqq}} \nn \\
 		\quad & -\frac{(\rC \dA2+\dA3) (\alVL-\alVR) \beVR \sqrt{w^2 - 1} \WpmP}{2 \sqrt{\mSqq}} \nn \\
 		\quad & -\frac{4 \dT1 \alTL \beTR \rt \sqrt{\Wm}}{\sqrt{\mSqq}} \nn \\
 		\quad & +\frac{2 \alTL \beTR \rt \big[-(\dT2+\dT3) \WpmP+\dT4 \sqrt{w^2 - 1} \WpmM\big]}{\sqrt{\mSqq}}\bigg\}\\
& A_{\dn\,1\,2\,\up}  = \bigg\{\frac{1}{2} \beSR (-\dP (\alSL-\alSR) \WpmP+\dS (\alSL+\alSR) \WpmM) \nn \\
 		\quad & -\frac{\Rp \dV1 (\alVL+\alVR) \beVR \rt}{2 \mSqq} \nn \\
 		\quad & -\frac{(\alVL+\alVR) \beVR \rt \WpmM (\dV3 \Op+\dV2 \Ot)}{2 \mSqq} \nn \\
 		\quad & +\frac{\Rm \dA1 (-\alVL+\alVR) \beVR \rt}{2 \mSqq} \nn \\
 		\quad & -\frac{(\alVL-\alVR) \beVR \rt \WpmP (\dA3 \Op+\dA2 \Ot)}{2 \mSqq} \nn \\
 		\quad & -4 \dT1 \alTL \beTR \sqrt{\Wm} \cos\thtau \nn \\
 		\quad & +2 \alTL \beTR \big[-(\dT2+\dT3) \WpmP+\dT4 \sqrt{w^2 - 1} \WpmM\big] \cos\thtau\bigg\}\\
& A_{\dn\,2\,1\,\dn}  = \sin\thtau\bigg\{-\frac{\dV1 (1+(\alVL+\alVR) \beVL) \rt \WpmP}{2 \sqrt{\mSqq}} \nn \\
 		\quad & -\frac{\dA1 (1+(\alVL-\alVR) \beVL) \rt \WpmM}{2 \sqrt{\mSqq}} \nn \\
 		\quad & +\frac{4 \dT1 \alTR \beTL (\Wm - \rC)}{\sqrt{\mSqq} \sqrt{\Wm}} \nn \\
 		\quad & +\frac{2 \alTR \beTL (\dT2-\rC \dT2 \Wp+\dT3 (\Wm - \rC)) \WpmP}{\sqrt{\mSqq}}\bigg\}\\
& A_{\dn\,2\,2\,\dn}  = \cos^2 \frac{\thtau}{2}\bigg\{-\dV1 (1+(\alVL+\alVR) \beVL) \WpmP \nn \\
 		\quad & -\dA1 (1+(\alVL-\alVR) \beVL) \WpmM \nn \\
 		\quad & -\frac{8 \dT1 \alTR \beTL \rt (\rC\Wp - 1) \sqrt{\Wm}}{\mSqq} \nn \\
 		\quad & +\frac{4 \alTR \beTL \rt (\dT2-\rC \dT2 \Wp+\dT3 (\Wm - \rC)) \WpmP}{\mSqq}\bigg\}\\
& A_{\dn\,2\,1\,\up}  = \sin^2 \frac{\thtau}{2}\bigg\{\dV1 (\alVL+\alVR) \beVR \WpmP \nn \\
 		\quad & +\dA1 (\alVL-\alVR) \beVR \WpmM \nn \\
 		\quad & +\frac{8 \dT1 \alTL \beTR \rt \sqrt{\Wp} (\rC\Wm - 1)}{\mSqq} \nn \\
 		\quad & +\frac{4 \alTL \beTR \rt (\dT3 (\rC-\Wp)+\dT2 (\rC\Wm - 1)) \WpmP}{\mSqq}\bigg\}\,.
\end{align}
\end{subequations}
Here we have written only the $s_b = \dn$ amplitudes. 
The $s_b = \up$ amplitudes for $\Lb \to \LcSp l \nu$ follow immediately from the conjugation relation
\begin{multline}
	A_{\bar{s}_b \bar{s}_c s_l s_\nu}(\thtau,w_+, w_-) \\
	\equiv h^l_{s_l s_\nu}(\pi)\, A_{s_b s_c  s_l s_\nu}(\pi-\thtau,w_-, w_+)\,,
\end{multline}
where $h^l$ is defined as in Eq.~\eqref{eqn:hdefs}, and the exchange $w_+ \leftrightarrow w_-$ implies $\sqrt{w^2 -1} \to -\sqrt{w^2-1}$ everywhere in the amplitudes.

Similarly, with respect to the total differential rate~\eqref{eqn:totdiff}, the explicit amplitudes for $\Lb \to \LcTn l \nu$ are then
\begin{widetext}
\begin{subequations}
\begin{align}
A_{\dn\,-\frac{1}{2}\,1\,\dn} & = \sin\thtau\bigg\{-\frac{\lV1 (1+(\alVL+\alVR) \beVL) \rt \sqrt{w^2 - 1} \WpmM}{\sqrt{6} \sqrt{\mSqq}} +\frac{\lV4 (1+(\alVL+\alVR) \beVL) \rt \WpmP}{2 \sqrt{6} \sqrt{\mSqq}} \nn \\
 		& +\frac{\lA1 (-1-(\alVL-\alVR) \beVL) \rt \sqrt{w^2 - 1} \WpmP}{\sqrt{6} \sqrt{\mSqq}}  +\frac{\lA4 (1+(\alVL-\alVR) \beVL) \rt \WpmM}{2 \sqrt{6} \sqrt{\mSqq}} \nn \\
 		& +\frac{2 \sqrt{\frac{2}{3}} \alTR \beTL \sqrt{w^2 - 1} \big[2 \lT1 (\rC\Wp - 1) \sqrt{\Wm}+(\lT2 (\rC\Wp - 1)+\lT3 (\rC-\Wm)) \WpmM\big]}{\sqrt{\mSqq}} \nn \\
 		& +\frac{\sqrt{\frac{2}{3}} \alTR \beTL \big[2 \lT4 (\rC-\Wm) \sqrt{\Wm}+(\lT5 (\rC\Wp - 1)+\lT6 (\rC-\Wm)) \WpmP\big]}{\sqrt{\mSqq}}\bigg\}\\
A_{\dn\,-\frac{1}{2}\,2\,\dn} & = \cos^2 \frac{\thtau}{2}\bigg\{-\sqrt{\frac{2}{3}} \lV1 (1+(\alVL+\alVR) \beVL) \sqrt{w^2 - 1} \WpmM  +\frac{\lV4 (1+(\alVL+\alVR) \beVL) \WpmP}{\sqrt{6}} \nn \\
 		& +\sqrt{\frac{2}{3}} \lA1 (-1-(\alVL-\alVR) \beVL) \sqrt{w^2 - 1} \WpmP  +\frac{\lA4 (1+(\alVL-\alVR) \beVL) \WpmM}{\sqrt{6}} \nn \\
 		& +\frac{4 \sqrt{\frac{2}{3}} \alTR \beTL \rt \sqrt{w^2 - 1} \big[2 \lT1 (\rC\Wp - 1) \sqrt{\Wm}+(\lT2 (\rC\Wp - 1)+\lT3 (\rC-\Wm)) \WpmM\big]}{\mSqq} \nn \\
 		& +\frac{2 \sqrt{\frac{2}{3}} \alTR \beTL \rt \big[2 \lT4 (\rC-\Wm) \sqrt{\Wm}+(\lT5 (\rC\Wp - 1)+\lT6 (\rC-\Wm)) \WpmP\big]}{\mSqq}\bigg\}\\
A_{\dn\,-\frac{1}{2}\,1\,\up} & = \sin^2 \frac{\thtau}{2}\bigg\{\sqrt{\frac{2}{3}} \lV1 (\alVL+\alVR) \beVR \sqrt{w^2 - 1} \WpmM  -\frac{\lV4 (\alVL+\alVR) \beVR \WpmP}{\sqrt{6}} \nn \\
 		& +\sqrt{\frac{2}{3}} \lA1 (\alVL-\alVR) \beVR \sqrt{w^2 - 1} \WpmP +\frac{\lA4 (-\alVL+\alVR) \beVR \WpmM}{\sqrt{6}} \nn \\
 		& +\frac{4 \sqrt{\frac{2}{3}} \alTL \beTR \rt \sqrt{w^2 - 1} \big[2 \lT1 \sqrt{\Wp} (\rC\Wm - 1)+(\lT2+\lT3 (\Wp - \rC)-\rC \lT2 \Wm) \WpmM\big]}{\mSqq} \nn \\
 		& +\frac{2 \sqrt{\frac{2}{3}} \alTL \beTR \rt \big[2 \lT4 \sqrt{\Wp} (\Wp - \rC)+(\lT5+\lT6 (\Wp - \rC)-\rC \lT5 \Wm) \WpmP\big]}{\mSqq}\bigg\}\\
A_{\dn\,-\frac{1}{2}\,2\,\up} & = \sin\thtau\bigg\{\frac{\lV1 (\alVL+\alVR) \beVR \rt \sqrt{w^2 - 1} \WpmM}{\sqrt{6} \sqrt{\mSqq}} -\frac{\lV4 (\alVL+\alVR) \beVR \rt \WpmP}{2 \sqrt{6} \sqrt{\mSqq}} \nn \\
 		& +\frac{\lA1 (\alVL-\alVR) \beVR \rt \sqrt{w^2 - 1} \WpmP}{\sqrt{6} \sqrt{\mSqq}} +\frac{\lA4 (-\alVL+\alVR) \beVR \rt \WpmM}{2 \sqrt{6} \sqrt{\mSqq}} \nn \\
 		& +\frac{2 \sqrt{\frac{2}{3}} \alTL \beTR \sqrt{w^2 - 1} \big[2 \lT1 \sqrt{\Wp} (\rC\Wm - 1)+(\lT2+\lT3 (\Wp - \rC)-\rC \lT2 \Wm) \WpmM\big]}{\sqrt{\mSqq}} \nn \\
 		& +\frac{\sqrt{\frac{2}{3}} \alTL \beTR \big[2 \lT4 \sqrt{\Wp} (\Wp - \rC)+(\lT5+\lT6 (\Wp - \rC)-\rC \lT5 \Wm) \WpmP\big]}{\sqrt{\mSqq}}\bigg\}\\
A_{\dn\,+\frac{1}{2}\,1\,\dn} & = \bigg\{\frac{\beSL \sqrt{w^2 - 1} (\lS \alSL \WpmP+\lS \alSR \WpmP-\lP \alSL \WpmM+\lP \alSR \WpmM)}{\sqrt{6}} \nn \\
 		& +\frac{\Rm \lV1 (1+(\alVL+\alVR) \beVL) \rt \sqrt{w^2 - 1}}{\sqrt{6} \mSqq} \nn \\
 		& -\frac{(1+(\alVL+\alVR) \beVL) \rt \WpmP \big[-\sqrt{w^2 - 1} (\lV3 \Op+\lV2 \Ot)+\lV4 \Oz\big]}{\sqrt{6} \mSqq} \nn \\
 		& +\frac{\Rp \lA1 (1+(\alVL-\alVR) \beVL) \rt \sqrt{w^2 - 1}}{\sqrt{6} \mSqq} \nn \\
 		& -\frac{(1+(\alVL-\alVR) \beVL) \rt \WpmM \big[-\sqrt{w^2 - 1} (\lA3 \Op+\lA2 \Ot)+\lA4 \Oz\big]}{\sqrt{6} \mSqq} \nn \\
 		& -2 \sqrt{\frac{2}{3}} \alTR \beTL \sqrt{w^2 - 1} \big[2 \lT1 \sqrt{\Wp}+(-\lT2+\lT3) \WpmM\big] \cos\thtau \nn \\
 		& -2 \sqrt{\frac{2}{3}} \alTR \beTL \big[2 \lT4 \sqrt{\Wm}+(\lT6+\lT5 w) \WpmP\big] \cos\thtau +2 \sqrt{\frac{2}{3}} \lT7 \alTR \beTL (w^2 - 1) \WpmP \cos\thtau\bigg\}\\
A_{\dn\,+\frac{1}{2}\,2\,\dn} & = \sin\thtau\bigg\{-\frac{(1+\rC) \lV1 (1+(\alVL+\alVR) \beVL) \sqrt{w^2 - 1} \WpmM}{\sqrt{6} \sqrt{\mSqq}} \nn \\
 		& +\frac{(1+(\alVL+\alVR) \beVL) \big[\lV3+\lV4 (\rC-w)-\lV3 w^2+\lV2 \big[\rC-\rC w^2\big]\big] \WpmP}{\sqrt{6} \sqrt{\mSqq}} \nn \\
 		& -\frac{(\rC - 1) \lA1 (1+(\alVL-\alVR) \beVL) \sqrt{w^2 - 1} \WpmP}{\sqrt{6} \sqrt{\mSqq}} \nn \\
 		& +\frac{(1+(\alVL-\alVR) \beVL) \big[\lA3+\lA4 (\rC-w)-\lA3 w^2+\lA2 \big[\rC-\rC w^2\big]\big] \WpmM}{\sqrt{6} \sqrt{\mSqq}} \nn \\
 		& +\frac{2 \sqrt{\frac{2}{3}} \alTR \beTL \rt \sqrt{w^2 - 1} \big[2 \lT1 \sqrt{\Wp}+(-\lT2+\lT3) \WpmM\big]}{\sqrt{\mSqq}} \nn \\
 		& +\frac{2 \sqrt{\frac{2}{3}} \alTR \beTL \rt \big[2 \lT4 \sqrt{\Wm}+(\lT6+\lT5 w) \WpmP\big]}{\sqrt{\mSqq}} -\frac{2 \sqrt{\frac{2}{3}} \lT7 \alTR \beTL \rt (w^2 - 1) \WpmP}{\sqrt{\mSqq}}\bigg\}\\
A_{\dn\,+\frac{1}{2}\,1\,\up} & = \sin\thtau\bigg\{-\frac{(1+\rC) \lV1 (\alVL+\alVR) \beVR \sqrt{w^2 - 1} \WpmM}{\sqrt{6} \sqrt{\mSqq}} \nn \\
 		& +\frac{(\alVL+\alVR) \beVR \big[\lV3+\lV4 (\rC-w)-\lV3 w^2+\lV2 \big[\rC-\rC w^2\big]\big] \WpmP}{\sqrt{6} \sqrt{\mSqq}} \nn \\
 		& -\frac{(\rC - 1) \lA1 (\alVL-\alVR) \beVR \sqrt{w^2 - 1} \WpmP}{\sqrt{6} \sqrt{\mSqq}} \nn \\
 		& +\frac{(\alVL-\alVR) \beVR \big[\lA3+\lA4 (\rC-w)-\lA3 w^2+\lA2 \big[\rC-\rC w^2\big]\big] \WpmM}{\sqrt{6} \sqrt{\mSqq}} \nn \\
 		& -\frac{2 \sqrt{\frac{2}{3}} \alTL \beTR \rt \sqrt{w^2 - 1} \big[2 \lT1 \sqrt{\Wm}+(\lT2-\lT3) \WpmM\big]}{\sqrt{\mSqq}} \nn \\
 		& +\frac{2 \sqrt{\frac{2}{3}} \alTL \beTR \rt \big[2 \lT4 \sqrt{\Wp}+(\lT6+\lT5 w) \WpmP\big]}{\sqrt{\mSqq}} -\frac{2 \sqrt{\frac{2}{3}} \lT7 \alTL \beTR \rt (w^2 - 1) \WpmP}{\sqrt{\mSqq}}\bigg\}\\
A_{\dn\,+\frac{1}{2}\,2\,\up} & = \bigg\{\frac{\beSR \sqrt{w^2 - 1} (\alSL (-\lS \WpmP+\lP \WpmM)-\alSR (\lS \WpmP+\lP \WpmM))}{\sqrt{6}} \nn \\
 		& -\frac{\Rm \lV1 (\alVL+\alVR) \beVR \rt \sqrt{w^2 - 1}}{\sqrt{6} \mSqq} \nn \\
 		& +\frac{(\alVL+\alVR) \beVR \rt \WpmP \big[-\sqrt{w^2 - 1} (\lV3 \Op+\lV2 \Ot)+\lV4 \Oz\big]}{\sqrt{6} \mSqq} \nn \\
 		& +\frac{\Rp \lA1 (-\alVL+\alVR) \beVR \rt \sqrt{w^2 - 1}}{\sqrt{6} \mSqq} \nn \\
 		& +\frac{(\alVL-\alVR) \beVR \rt \WpmM \big[-\sqrt{w^2 - 1} (\lA3 \Op+\lA2 \Ot)+\lA4 \Oz\big]}{\sqrt{6} \mSqq} \nn \\
 		& -2 \sqrt{\frac{2}{3}} \alTL \beTR \sqrt{w^2 - 1} \big[2 \lT1 \sqrt{\Wm}+(\lT2-\lT3) \WpmM\big] \cos\thtau \nn \\
 		& +2 \sqrt{\frac{2}{3}} \alTL \beTR \big[2 \lT4 \sqrt{\Wp}+(\lT6+\lT5 w) \WpmP\big] \cos\thtau -2 \sqrt{\frac{2}{3}} \lT7 \alTL \beTR (w^2 - 1) \WpmP \cos\thtau\bigg\}\\
A_{\dn\,+\frac{3}{2}\,1\,\dn} & = \sin\thtau\bigg\{-\frac{\lV4 (1+(\alVL+\alVR) \beVL) \rt \WpmP}{2 \sqrt{2} \sqrt{\mSqq}}  -\frac{\lA4 (1+(\alVL-\alVR) \beVL) \rt \WpmM}{2 \sqrt{2} \sqrt{\mSqq}} \nn \\
 		& -\frac{\sqrt{2} \alTR \beTL \big[2 \lT4 \sqrt{\Wp} (\rC\Wm - 1)+(\lT6 (\rC-\Wp)+\lT5 (\rC\Wm - 1)) \WpmP\big]}{\sqrt{\mSqq}}\bigg\}\\
A_{\dn\,+\frac{3}{2}\,2\,\dn} & = \sin^2 \frac{\thtau}{2}\bigg\{\frac{\lV4 (1+(\alVL+\alVR) \beVL) \WpmP}{\sqrt{2}} +\frac{\lA4 (1+(\alVL-\alVR) \beVL) \WpmM}{\sqrt{2}} \nn \\
 		& +\frac{2 \sqrt{2} \alTR \beTL \rt \big[2 \lT4 \sqrt{\Wp} (\rC\Wm - 1)+(\lT6 (\rC-\Wp)+\lT5 (\rC\Wm - 1)) \WpmP\big]}{\mSqq}\bigg\}\\
A_{\dn\,+\frac{3}{2}\,1\,\up} & = \cos^2 \frac{\thtau}{2}\bigg\{-\frac{\lV4 (\alVL+\alVR) \beVR \WpmP}{\sqrt{2}} +\frac{\lA4 (-\alVL+\alVR) \beVR \WpmM}{\sqrt{2}} \nn \\
 		& -\frac{2 \sqrt{2} \alTL \beTR \rt \big[2 \lT4 (\rC\Wp - 1) \sqrt{\Wm}+(\lT5 (\rC\Wp - 1)+\lT6 (\rC-\Wm)) \WpmP\big]}{\mSqq}\bigg\}\,.
\end{align}
\end{subequations}
\end{widetext}
Here, again, we have written only the $s_b = \dn$ amplitudes.
The $s_b = \up$ amplitudes for $\Lb \to \LcTn l \nu$ follow from the conjugation relation
\begin{align}
	&A_{\bar{s}_b \bar{s}_c s_l s_\nu}(\thtau,w_+, w_-) \\
	& \qquad \equiv e^{i (|s_c| + 1/2) \pi} h^l_{s_l s_\nu}(\pi)\, A_{s_b  s_c s_l s_\nu}(\pi-\thtau,w_-, w_+)\,, \nn
\end{align}
where $h^l$ is defined as in Eq.~\eqref{eqn:hdefs}, and the exchange $w_+ \leftrightarrow w_-$ implies $\sqrt{w^2 -1} \to -\sqrt{w^2-1}$ everywhere in the amplitudes.

\section{Form Factor Bases}
\label{app:FFbasis}
The transformation from the LQCD helicity basis~\cite{Meinel:2021rbm} to the HQET basis~\eqref{eqn:HQETffdefSp} for $\Lb \to \LcSp$
\begin{subequations}
\begin{align}
f_{0} & = \dV1 + \frac{(\rC w - 1)\dV2}{1 + \rC } + \frac{(\rC  - w)\dV3}{1 + \rC }\,, \\
f_{+} & = \dV1 - \frac{(w - 1)[\rC \dV2+ \dV3]}{1 - \rC }\,, \\
f_{\perp} & = \dV1\,, \\
g_{0} & = \dA1 - \frac{(\rC w - 1)\dA2}{1 - \rC } - \frac{(\rC  - w)\dA3}{1 - \rC }\,, \\
g_{+} & = \dA1 + \frac{ (w + 1)[\rC \dA2 + \dA3]}{1 + \rC }\,, \\
g_{\perp} & = \dA1\,, \\
h_{+} & = \dT1 + \dT2 + \dT3 + (1 - w)\dT4\,, \\
h_{\perp} & = \dT1 - \frac{(\rC w - 1)\dT2}{1 - \rC } - \frac{(\rC  - w)\dT3}{1 - \rC }\,, \\
\tilde{h}_{+} & = \dT1\,, \\
\tilde{h}_{\perp} & = \dT1 + \frac{(w + 1)[\rC \dT2 + \dT3]}{1 + \rC }\,,
\end{align}
\end{subequations}
which are denoted with a ``$(\frac12^-)$'' superscript in Ref.~\cite{Meinel:2021rbm}.

The transformation from the LQCD helicity basis~\cite{Descotes-Genon:2019dbw,Meinel:2020owd,Meinel:2021rbm} to the HQET basis in Eqs.~\eqref{eqn:HQETffdefTn} and~\eqref{eqn:HQETffdefTnT} for $\Lb \to \LcTn$
\begin{subequations}
\begin{align}
	F_{0} & = 2(w + 1)\lV1 - \frac{2(w + 1)(\rC w - 1)\lV2}{1 - \rC } \nn\\
		& \quad - \frac{2(\rC  - w)(w + 1)\lV3}{1 - \rC } + \frac{2(w + 1)\lV4}{1 - \rC }\,, \\
	F_{+} & = 2(w - 1)\lV1 + \frac{2 (w^2 -1)[\rC \lV2 + \lV3]}{1 + \rC }  \nn\\
		& \quad - \frac{2(\rC  - w)\lV4}{1 + \rC }\,, \\
	F_{\perp} & = 2(w - 1)\lV1 - \lV4\,, \\
	F_{\perp^\prime} & = \lV4\,, \\
	G_{0} & = 2(w - 1)\lA1 + \frac{2(w - 1)(\rC w - 1)\lA2}{1 + \rC } \nn\\
		& \quad + \frac{2(\rC  - w)(w - 1)\lA3}{1 + \rC } - \frac{2(w - 1)\lA4}{1 + \rC }\,, \\
	G_{+} & = 2(w + 1)\lA1 - \frac{2 (w^2 -1)[\rC\lA2 + \lA3]}{1 - \rC } \nn\\
		& \quad + \frac{2(\rC  - w)\lA4}{1 - \rC }\,, \\
	G_{\perp} & = 2(w + 1)\lA1 - \lA4\,, \\
	G_{\perp^\prime} & = \lA4\,, \\
	H_{+} & = 2(w - 1)[\lT1 - \lT2 + \lT3] \nn\\
		& \quad + 2\lT4 + 2w\lT5 + 2\lT6 -2 (w^2 - 1)\lT7\,, \\
	H_{\perp} & = 2(w - 1)\lT1 + \frac{2(w - 1)(\rC w - 1)\lT2}{1 + \rC } \nn\\
		& \quad + \frac{2(\rC  - w)(w - 1)\lT3}{1 + \rC } + \frac{(1 + \rC  - 2w)\lT4}{1 + \rC } \nn\\
		& \quad +  \frac{(\rC w - 1)\lT5}{1 + \rC } + \frac{(\rC  - w)\lT6}{1 + \rC }\,, \\
	H_{\perp^\prime} & = \frac{(1 - \rC )\lT4}{1 + \rC } + \frac{(1 - \rC w)\lT5}{1 + \rC } \nn\\
		& \quad + \frac{(w -\rC)\lT6}{1 + \rC }\,, \\
	\tilde{H}_{+} & = 2(w + 1)\lT1 - 2\lT4\,, \\
	\tilde{H}_{\perp} & = 2(w + 1)\lT1 - \frac{2 (w^2 -1)[\rC \lT2 + \lT3]}{1 - \rC } \\
		& \quad - \frac{(1 - \rC  + 2w)\lT4}{1 - \rC } -  \frac{(w + 1)[\rC \lT5 + \lT6]}{1 - \rC } \,\nn\\
	\tilde{H}_{\perp^\prime} & = -\frac{(1 + \rC )\lT4}{1 - \rC } - \frac{(w + 1)[\rC\lT5 + \lT6]}{1 - \rC } \,.
\end{align}
\end{subequations}
In Ref.~\cite{Meinel:2021rbm} each of these $\Lb \to \LcTn$ form factors is denoted with a corresponding lower case base symbol, and a ``$(\frac32^-)$'' superscript.
One may verify straightforwardly that the six tensor form factors are orthogonal to the kernel~\eqref{eqn:lTredef}.

\section{LQCD fit results}
\label{app:nuisance}

Performing the LQCD form factor fit with HQS-breaking parameters as described in Sect.~\ref{sec:LQCD}, one obtains
\begin{align}
  \sigma(1) &=  0.9\pm 0.2, &\sigma' &= -2.2\pm 0.6, \nn\\ 
  \hs(1) &=  0.6\pm 0.2, & \nn\\
  \hpc(1) &=  -0.8\pm 0.5, &\hpb(1) &= -0.2\pm 0.3,  \nn\\
  \epsilon_{\dV1} &=  0.01\pm 0.03, &\epsilon_{\dV2} &= 0.7\pm 0.2, \nn\\ 
  \epsilon_{\dV3} &= 0.06\pm 0.07,  &\epsilon_{\dA1} &=  -0.8\pm 0.2\nn\\
  \epsilon_{\dA2} &= 0.8\pm 0.2,  &\epsilon_{\dA3} &= -0.05\pm 0.02,  \nn\\
  \epsilon_{\dT1} &=  -0.9\pm 0.2,   &\epsilon_{\dT2} &= 0.9\pm 0.2, \nn\\ 
  \epsilon_{\dT3} &= -0.01\pm 0.02,  &\epsilon_{\dT4} &= 0.04\pm 0.098,  \nn\\
  \epsilon_{\lV1} &=  0.1\pm 0.3,   &\epsilon_{\lV2} &= 0.2\pm 0.2, \nn\\ 
  \epsilon_{\lV3} &= -0.20\pm 0.05,    &\epsilon_{\lV4} &= 0.01\pm 0.01,  \nn\\
  \epsilon_{\lA1} &=  0.0\pm 0.3,   &\epsilon_{\lA2}& = 0.1\pm 0.2, \nn\\ 
  \epsilon_{\lA3} &= 0.3\pm 0.1,  &\epsilon_{\lA4} &= -0.04\pm 0.08,  \nn\\
  \epsilon_{\lT1} &=  0.0\pm 0.3, &\epsilon_{\lT2} &= 0.3\pm 0.3, \nn\\ 
  \epsilon_{\lT3} &= 0.1\pm 0.1,  &\epsilon_{\lT4} &= 0.01\pm 0.10,  \nn\\
  \epsilon_{\lT6} &=  -0.01\pm 0.08, &\epsilon_{\lT7} &=-0.1 \pm 0.2\,.
\end{align}
The uncertainties are estimated using MC sampling of the likelihood function, 
to generate a profile likelihood for each parameter and determine its values at which $\Delta \chi^2 = 1$.

\section{Amplitude counting and tensor polarizations with the spinor helicity formalism}
\label{app:helicity}

Here we use the formalism of Ref.~\cite{Arkani-Hamed:2017jhn}. 
Since insertions of $p_{\alpha \dot\alpha}/m$ can always be used for massive particles to convert dotted to undotted indices, 
we can write the amplitude---stripped of the external massive spinor helicity variables---as a tensor in the undotted indices, 
and totally symmetric in the each of the $2S_i$ indices associated to the external particle carrying spin $S_i$. 
Such a tensor can be expressed in terms of products of $O_{\alpha\beta}=p_{1\{\alpha\dot\beta}p_{2\beta\}}^{\dot\beta}$ and (at most one) $\epsilon_{\alpha\beta}$ factor. 
In the case of $\Lb\to\LcTn$ mediated by the (pseudo)tensor operator, 
one has a 3-point amplitude between a spin-$1/2$, spin-$3/2$ and a spin-$1$ particle, since $J^P(\mathcal{O}_T) = 1^+ \oplus 1^-$.
Therefore, counting all the possible independent amplitudes is equivalent to counting how many ways one can construct 
$M_{\alpha_1,\beta_1\beta_2,\gamma_1\gamma_2\gamma_3}$ (symmetric in the $\beta$ and $\gamma$ indices) 
in terms of either: (a) three $O_{\alpha\beta}$'s; or (b) two $O_{\alpha\beta}$'s and one $\epsilon_{\alpha\beta}$. 
It is straightforward to see that the number of possible contractions is $3+3 = 6$.

Considering insertions of $p_{\alpha\dot\alpha}/m$, one may also easily see how the different representations of the spin-$1$ polarization tensors are equivalent. 
The canonical representation for $\epsilon_{\alpha\beta}^{0,+,-}$ polarization vectors is
\begin{equation}
	\epsilon_{\alpha\beta}^-=\frac{\lambda_\alpha\lambda_\beta}{m}\,, \quad 
	\epsilon_{\alpha\beta}^0=\frac{\lambda_\alpha\eta_\beta + \eta_\alpha\lambda_\beta}{2m}\,, \quad  
	\epsilon_{\alpha\beta}^+=\frac{\eta_\alpha\eta_\beta}{m}\,,
\end{equation}
while $\epsilon_{\alpha\dot\alpha}^{0,+,-}$ is usually written as
\begin{equation}
	\epsilon_{\alpha\dot\alpha}^-=\frac{\lambda_\alpha\tilde\eta_{\dot\alpha}}{m}\,, \quad 
	\epsilon_{\alpha\dot\alpha}^0=\frac{\lambda_\alpha\tilde\lambda_{\dot\alpha} + \eta_\alpha\tilde\eta_{\dot\alpha}}{2m}\,, \quad 
	\epsilon_{\alpha\beta}^+=\frac{\eta_\alpha\tilde\lambda_{\dot\alpha}}{m}
\end{equation}
and the momentum $q_{\alpha\dot\alpha} = \lambda_\alpha \tilde\lambda_{\dot\alpha} + \eta_\alpha \tilde\eta_{\dot\alpha}$.
It is trivial to show that
\begin{equation}
	\epsilon^{[+}_{\alpha\dot\beta}\epsilon^{-]\dot\beta}_\gamma =
	\frac{1}{m^2}\left(\lambda_\alpha\tilde\eta_{\dot\beta}\tilde\lambda^{\dot\beta}\eta_\gamma-\eta_\alpha\tilde\lambda_{\dot\beta}\lambda_\alpha\tilde\eta^{\dot\beta}\right)= -\epsilon_{\alpha\gamma}^0
\end{equation}
where we have used $[\tilde\lambda\tilde\eta]=m$. 
The case of $\epsilon_{\alpha\dot\beta}^0 q^{\dot \beta}_\gamma$ is even more trivial, as contracting with $q^{\dot \beta}_\gamma / m$ 
converts $\epsilon_{\alpha\dot\beta}^0$ into $\epsilon_{\alpha\gamma}^0$.
Similar relations hold for the other polarizations. 
Returning to Lorentz index notation, one reproduces Eq.~\eqref{eqn:epsqidentity}.

\end{document}